\documentclass[twocolumn,secnumarabic,amssymb, nobibnotes, aps, prd]{revtex4-1}

\usepackage{graphicx,bm,lmodern,pdftexcmds,amsmath,stmaryrd,siunitx,amssymb,xcolor}

\DeclareMathAlphabet{\mathsfbi}{OT1}{\sfdefault}{bx}{sl}
\DeclareMathVersion{sfletters}
\SetSymbolFont{letters}{sfletters}{OML}{ntxsfmi}{b}{it}

\newcommand{\bcdot}{\boldsymbol{\cdot}}
\newcommand{\bnabla}{\boldsymbol{\nabla}}

\begin{document}

\title{Computing the motor torque of \textit{Escherichia coli}}%

\author{Debasish Das and Eric Lauga}%
\email[]{E-mail: e.lauga@damtp.cam.ac.uk}
\affiliation{\textit{Department of Applied Mathematics and Theoretical Physics, Centre for Mathematical Sciences, University of Cambridge, Wilberforce Road, Cambridge CB3 0WA, UK.}}
\date{June 16, 2018}%

\begin{abstract}
The rotary motor of  bacteria is a natural nano-technological marvel that enables cell  locomotion by powering the rotation of  semi-rigid helical  flagellar filaments in  fluid environments. It is well known that the motor operates essentially at constant torque in counter-clockwise direction but past work have reported a large range of values of this torque. 
Focusing on \textit{Escherichia coli} cells that are swimming and cells that are stuck on a glass surface for which all geometrical and environmental parameters are known (N. C. Darnton et al., \textit{J.~Bacteriology}, 2007, \textbf{189}, 1756--1764), we use two validated numerical methods to compute the value of the motor torque consistent with   experiments. 
Specifically, we use (and compare) a numerical method based  on the boundary integral representation of Stokes flow   and also develop a hybrid method combining boundary element and slender body theory to model the cell body and flagellar filament, respectively. 
Using measured rotation speed of the motor, our computations predict a value of the motor torque in the  range  440 pN~nm to 829 pN~nm, depending critically on the distance between the flagellar filaments and the nearby surface. 
\end{abstract}

\maketitle
\section{Introduction}\label{sec:intro}
The study of bacteria is not only important in order to understand many pathogenic diseases, it also serves as a  model system for microorganisms locomotion \cite{berg2008,lauga2009,lauga2016} and their  response to environmental cues \cite{berg1975,ottemann1997}. Bacteria are present in great abundance on Earth in all kinds of environments ranging from living creatures such as plants and animals to non-living bodies like soil\cite{gans2005}, rock\cite{gorbushina2009}, oceans\cite{azam1983} and  sea ice \cite{staley1999}. 

Bacteria have devised  different techniques to perform locomotion contingent on their surrounding. As many as six different types of translocation have been reported in literature \cite{henrichsen1972,harshey2003}, namely (i) swarming, requiring excessive development of flagella; (ii) swimming, depending crucially on interactions between flagella and fluid; (iii) gliding, dependent on an intrinsic motive force; (iv) twitching, using an intrinsic motive force on type IV pili (slender appendages that attach to substrates and pull on the cell body); (v) sliding, requiring  growth and reduced friction; and (vi) darting, dependent on growth of capsulated aggregates. 

In this article, we consider  the swimming mode  of bacteria locomotion and focus on the model organism \textit{Escherichia coli} (\textit{E.~coli}). A lot is known about this bacterium from the point of view of genetics and biochemistry, which enables researchers to  induce  behavioural changes and elucidate its motility mechanisms \cite{berg2008}. 

\begin{figure}[b]
\centering
\includegraphics[width=0.42\textwidth]{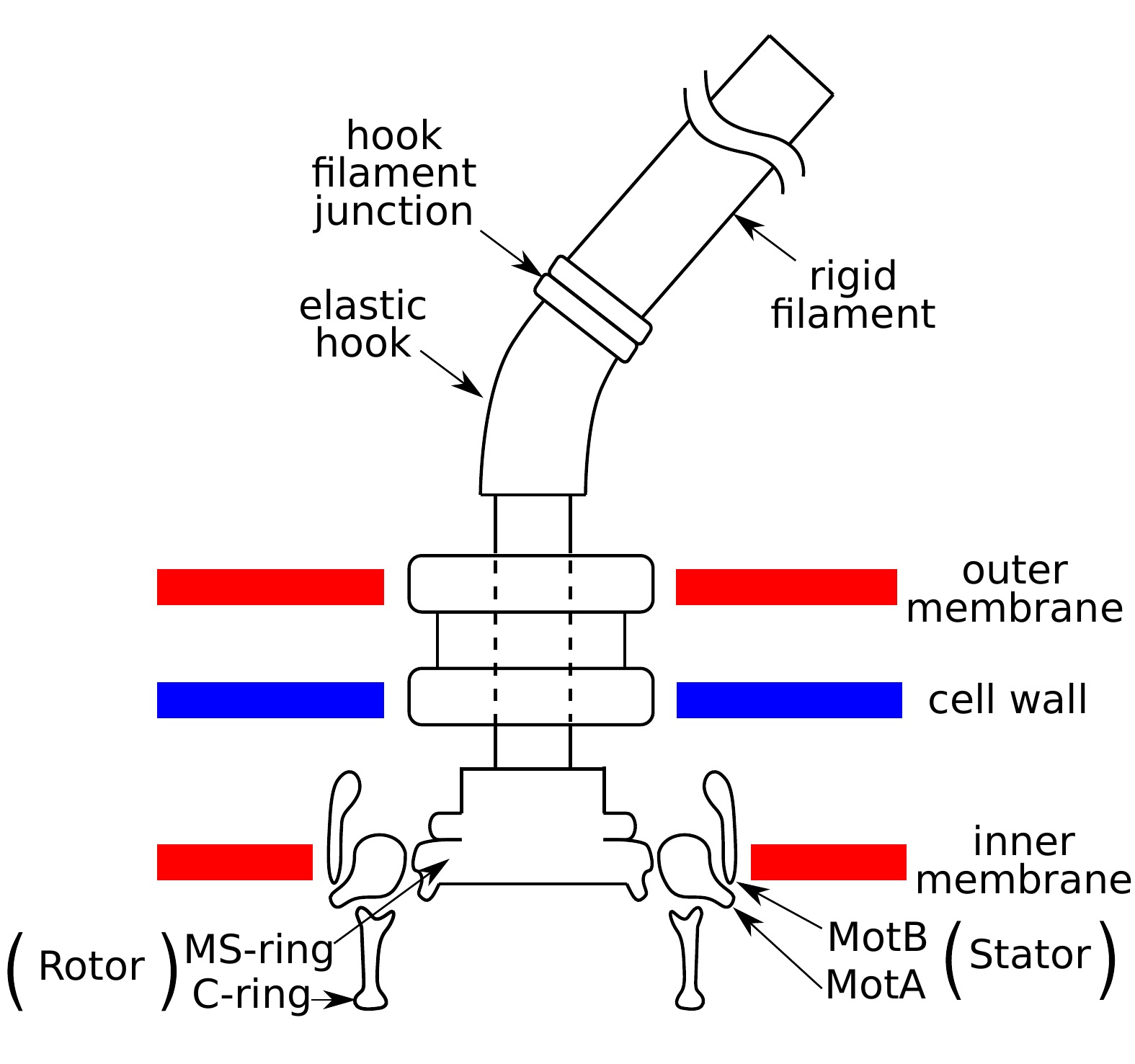}
\caption{Schematic diagram of the bacterial flagellar motor (adapted from Berg \cite{berg2003}). The primary components include a rigid filament attached to an elastic hook rotated by torque generating protein units MotA and MotB that act as stator while the MS-ring and C-ring form the rotor.}
\label{fig:motor}
\end{figure}

The locomotion of \textit{E.~coli} is powered by rotary motors rotating helical flagellar filaments, see Fig.~\ref{fig:motor}. Each motor is embedded in a 3-layered cell wall. Structures outside the cell wall include a rigid helical flagellum attached to the motor via a short elastic hook, while those inside include a basal body and a short rod.  
The rod acts as a drive shaft and rotates the elastic hook which plays the role of a universal joint, so that it can rotate the semi-rigid flagellar filament even in situations when the rod and filament are not aligned coaxially. \cite{berg2003,berg2008}.

Directional reversibility of the rotary motor is the main mechanism behind the ``run and tumble'' motion  of bacteria  searching for favourable environments\cite{berg1993a}. On an average, an \textit{E.~coli} cell has four such motors located randomly on its surface\cite{turner05}. During a run event, left-handed flagella arising from these motors form a helical bundle behind the cell and rotate counter-clockwise (CCW), when viewed from behind the cell, thereby pushing it forward. To initiate a tumble event, at least one of the motors starts to rotate clockwise (CW), the flagellum attached to this motor comes out of the bundle and transforms from a left-handed to a right-handed helical shape, still pushing the cell forward. This enables the cell to change its course of motion. After a short period,  the reversed motor switches back to CCW rotation and the flagellum returns to its normal left-handed configuration, rejoining the bundle\cite{turner05}. 
 
Torque in the motor is generated by interactions between the rotor (MS-ring and C-ring) and the stator (MotA and MotB proteins) units, the latter functioning as proton channels, see illustration in Fig.~\ref{fig:motor}. Experiments have reported at least 11 torque generating units, each consisting of a stator \cite{reid2006}. Proton flux through the proton channels due to an electrochemical gradient causes the rotor to rotate. The work done per unit charge that a proton does in crossing the cell membrane is called the protonmotive force (pmf). Though the motor's speed is known to be proportional to pmf, the exact mechanism of torque generation is still an active area of research \cite{gabel2003,berg2008}. Central to the work in our paper is the exact value of the  torque generated by the motor. A variety of experimental methods have been used to measure the torque-speed relationship for a number of bacteria species \cite{sowa2008}. For {\it E.~coli}, the torque is approximately constant up to a speed of a few 100 Hz, before rapidly decreasing to zero at a critical zero-torque speed.  

A wide range of values have been reported in the literature for the constant torque at frequencies below the constant speed. Early work indirectly measured the motor torque using (i) electrorotation \cite{berg1993b} where the cells are made to rotate using electric fields; (ii) tethered cell experiments \cite{berry1997} where the flagella is tethered to a glass surface and the cell body rotates about a fixed point;  and (iii) tethered bead experiments \cite{reid2006} where the flagellar filament is sheared off and a 
 spherical bead is attached to the remaining flagellar stub. The relatively slower rotation speeds and simpler spherical geometry make it easier to track the rotation speed of these beads, allowing to obtain  an estimation of the value of the motor torque. More recent experiments \cite{van2017} used 
  spherical magnetic beads to replace the filament and an external applied magnetic field to stop them from rotating, allowing measurement of the stall motor torque.

Alternatively, the  value of the motor torque  could be inferred directly by combining modelling with a  measure of the rotation speed of flagellar filaments of a swimming \textit{E. coli}.  Swimming  bacteria typically actuate independently  between four and seven  filaments which are gathered behind the cell in a thick helical  bundle \cite{turner05}, a geometrical setup which  presents modelling challenges due to filament-filament interactions \cite{flores2005study}. A clever alternative  was reported by  Darnton et al.~\cite{darnton2007} who, in addition to focusing on swimming cells,  considered  bacteria  stuck on a glass surface which only rotate a single flagellar filament. This simplified situation makes it  an ideal case to analyse from a modelling and numerical simulation viewpoint. 
 
In this paper, we combine the boundary element method and slender body theory in order to develop a mathematical model of the flagellated bacterium {\it E.~coli} and numerically simulate the experiments of Darnton et al.~\cite{darnton2007}. We first address the  case where the cell is freely swimming, despite some experimental unknowns, in order to compare the experimental observables with those obtained in the numerics.   We then consider the situation where the cell body is stuck on the wall, a configuration where all characteristics of the experiments are precisely known. Boundary element methods have long been used   to study   problems in bacteria locomotion including flagellar propulsion \cite{phan1987}, interaction between two swimming bacteria \cite{ishikawa2007}, bacterial behaviour and entrapment close to surfaces \cite{shum2010,giacche2010}, and locomotion using  multiple flagella \cite{kanehl2014}. Similarly, slender-body theory has been used to address problems in bacterial swimming such as bacterial polymorphism and optimal propulsion \cite{spagnolie2011} and microscale pumping by bacteria near walls \cite{dauparas2018}. In addition to our two numerical studies, we compare our results with two recent attempts to simulate the experiments of Darnton et al.~\cite{darnton2007} using mesoscale hydrodynamic simulations \cite{hu2015} and bead-spring model \cite{kong2015}.
 
Our paper is organised as follows. We describe the geometry and parameters of the problem in \S \ref{sec:geometry} and the basics of rigid body boundary element method in \S \ref{sec:boundaryelementmethod}. We next introduce the two computational approaches used in this study, based solely on boundary element method in \S \ref{sec:cm-I} and a hybrid method based on boundary element and slender body theory in \S \ref{sec:cm-II}. We validate the two models with existing semi-analytical results in \S \ref{sec:validation}. The results of our numerical simulations are presented in \S \ref{sec:results}, first for free-swimming cells, and then for cells stuck on surfaces, followed by a comparison with previous work in \S \ref{sec:comparison} and discussion in \S \ref{sec:discussion}. The results of various tests performed to validate the numerical method against tractable analytical solutions and details of mesh generation are provided in appendices.

%%%%%%%%%%%%%%%%%%%%%%%%%%%%
\section{Model: Geometry and computations}\label{sec:model}

\subsection{Geometry and parameters}\label{sec:geometry}

\begin{figure}
\centering
\includegraphics[width=0.5\textwidth]{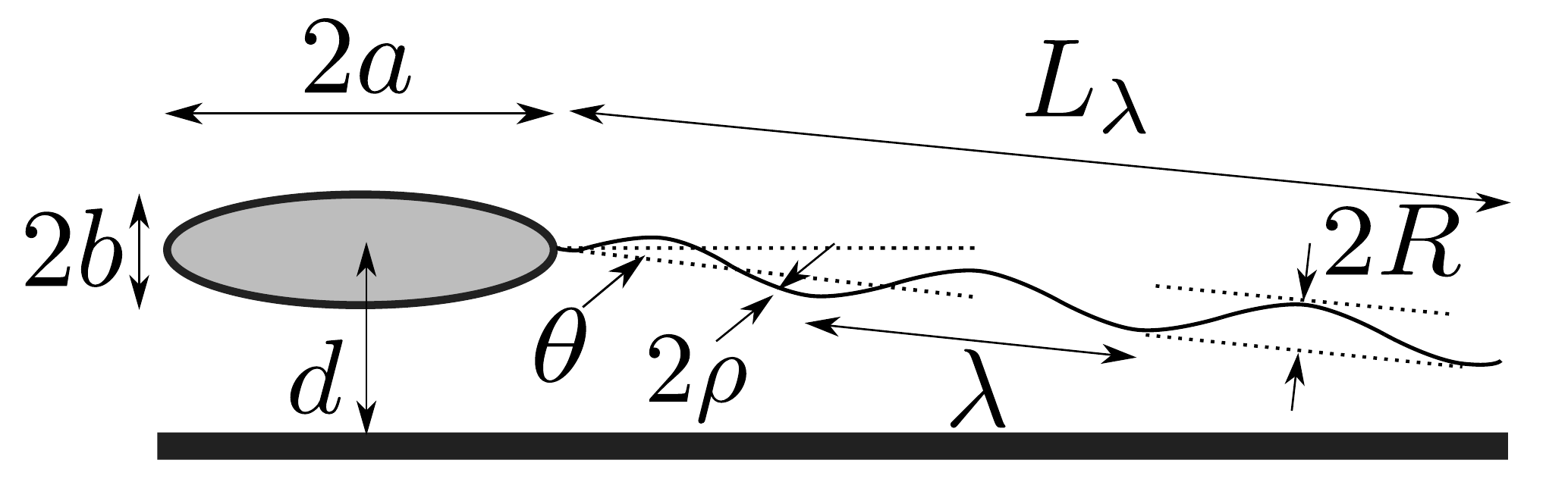}
\caption{Schematic diagram of a bacterium with a cell body modelled as a prolate spheroid of length $2a$, width $2b$ and a rigid tapered helical flagellum of pitch $\lambda$, radius $R$, axial length $L_\lambda$, filament cross-sectional radius $\rho$, placed at a distance $d$ parallel to a rigid wall and tilted at an angle $\theta$ with respect to the bottom surface.}
\label{fig:schematic}
\end{figure}

We use the geometrical dimensions of bacteria reported in the experiments of Darnton et al.~\cite{darnton2007}. The  cell body is \SI{2.5}{\micro\metre} long and \SI{0.88}{\micro\metre} wide (see Fig.~\ref{fig:schematic}). The length of the helical flagellar filament  measured along its centreline is $\sim$ \SI{7}{\micro\metre} and the radius of the helix is \SI{0.2}{\micro\metre}. The viscosity of the aqueous medium where they reside ranges from  $\mu =0.931$~cP to  $3.07$~cP. In the case of free-moving cells, the swimming speed of bacteria and rotational velocity of the flagellar filaments are measured to be $29$ \SI{}{\micro\metre}$/s$ and $2\pi \times 131$ Hz respectively. Taking the density of water to be $1000$ $\mathrm{kg/m}^3$, the corresponding Reynolds numbers for translation and rotation are $\sim 7\times10^{-5}$ and $\sim 3\times10^{-5}$, respectively, small enough  that fluid inertia can be neglected. This allows us to describe fluid motion around the bacteria using the incompressible Stokes equation, 
\begin{equation}\label{eq:stokes}
-\bnabla p + \mu \nabla^2 \bm{u} = 0,\quad \bnabla \bcdot \bm{u} = 0,
\end{equation}
where $p$ and $\bm{u}$ are the pressure and velocity of the fluid.

%%%%%%
\subsection{Boundary element method (BEM)}\label{sec:boundaryelementmethod}
We use  boundary element method to solve the flow problem around rigid bodies with prescribed motion\cite{pozrikidis1992,pozrikidis2002}. Consider a rigid body experiencing a hydrodynamic stress that resists its motion. The disturbance velocity field in a fluid domain $V$, free of any additional localised forcing, is represented by a surface distribution of flow singularities called stokeslets whose strength is the modified boundary traction, $\bm{f}_h$, i.e.~we have
\begin{equation}\label{eq:bemslp}
\bm{u}(\bm{x}_0) = \bm{u}_\infty(\bm{x}_0) - \frac{1}{8\pi\mu} \iint\displaylimits_{S} \bm{f}_h (\bm{x}) \bcdot \mathsfbi{G}(\bm{x},\bm{x}_0) \, \mathrm{d} S(\bm{x}).
\end{equation}
In Eq.~\eqref{eq:bemslp},  the integration points $\bm{x}$ are on $S$, the boundary of the fluid domain (i.e.~the surface of the body) and the evaluation point $\bm{x}_0$ is any point in $V$, the bulk fluid, and $\bm{u}_\infty$ is an arbitrary external flow set to zero here. The mathsfbi  $\mathsfbi{G}(\bm{x},\bm{x}_0)$ is the free space Green's function for Stokes equation, Eq.~\eqref{eq:stokes}, also called the Oseen-Burgers mathsfbi, representing fluid flow produced by a point force,
\begin{equation}\label{eq:oseenmathsfbi}
\mathsfbi{G}(\bm{x},\bm{x}_0) = \frac{\mathsfbi{I}}{|\bm{x}-\bm{x}_0|} + \frac{(\bm{x}-\bm{x}_0)(\bm{x}-\bm{x}_0)}{|\bm{x}-\bm{x}_0|^3},
\end{equation}
where $\mathsfbi{I}$ is the $3\times3$ identity mathsfbi. The integral equation relating the interfacial velocity and the surface traction via distribution of stokeslet is referred to as single layer potential (SLP) integral. When the evaluation point lies on the surface, $\bm{x}_0 \in S$, the no slip condition prescribes the fluid motion to be the same as the surface of the rigid body. The surface velocity of a body translating with velocity $\bm{U}$, measured at some arbitrary origin inside the body, and rotating with velocity $\bm{\Omega}$, measured about a point $\bm{x}_m$, is
\begin{equation}\label{eq:kinematics}
\bm{u}(\bm{x}_0) = \bm{U} + \bm{\Omega} \times (\bm{x}_0-\bm{x}_m), \quad \bm{x}_0 \in S.
\end{equation}
The surface of the rigid body is discretised into 6-nodes curved elements and each element is assumed to have a constant velocity and force. Though 3-nodes flat elements are applicable for this problem as well, curved elements are used for better accuracy. When the evaluation point lies on the surface, $\bm{x}_0 \in S$, the kernel in equation \eqref{eq:bemslp} can become singular and the singularity is removed by transforming the variables from Cartesian to polar coordinates\cite{pozrikidis2002}. For a prescribed velocity, $\bm{u}$, the unknown hydrodynamic surface tractions on the surface $\bm{f}_h$ can be computed by solving equation \eqref{eq:bemslp} numerically. The net hydrodynamic force, $\bm{F} $,  and torque, $\bm{T}$ measured about point $\bm{x}_m$, acting on the body are then computed as 
\begin{equation}\label{eq:netforce}
\begin{split}
\bm{F} &=\iint\displaylimits_S \bm{f}_h(\bm{x}) \,\mathrm{d}S(\bm{x}), \\
\bm{T} &=\iint\displaylimits_S (\bm{x}-\bm{x}_m) \times \bm{f}_h(\bm{x}) \,\mathrm{d}S(\bm{x}),  \quad \bm{x} \in S.
\end{split}
\end{equation}
The linear system arising from the SLP formulation is known for being ill-conditioned making it unsuitable for iterative solvers like GMRES \cite{saad1986} used in this work. However, in practice this is only a problem for very fine discretisation giving rise to large linear systems and the SLP formulation has been used with success in many previous studies involving swimming micro-organisms \cite{phan1987,ishikawa2007,smith2009,shum2010,giacche2010,kanehl2014}.  The fluid dynamics of a swimming bacterium involves solving the flow interaction between the cell body and the flagella. Since the flagella are slender, we have the option of discretising their entire surface and tackling it also using BEM, as carried out in \S \ref{sec:cm-I}, or taking advantage of slender body theories \cite{johnson1980} (SBT), thereby reducing the surface integrals on the filaments to contour integrals, as we do in \S \ref{sec:cm-II}.

\subsection{Computational model I: BEM-BEM}\label{sec:cm-I}

In this first computational model (termed CM-I), both the cell body and flagellum surfaces, denoted as  $S_b$ and $S_f$ respectively, are discretised as illustrated in Fig.~\ref{fig:backview}. The details of the surface discretisation method are provided in appendix \ref{sec:discretisation}. The cell body is modelled as a prolate spheroid with $x$-direction being the symmetry axis, of equation
\begin{equation}\label{eq:spheroid}
\frac{x^2}{a^2} + \frac{y^2 + z^2}{b^2} = 1,
\end{equation} 
where $a$ and $b$ are the major and minor semi-axes length respectively ($a>b$). The flagellum is modelled as a rigid left-handed helix with tapered ends \cite{higdon1979a,higdon1979b},
\begin{equation}
\begin{split}
\bm{x} = [s,E(s)R\sin(ks+\psi),E(s)R&\cos(ks+\psi)], \\
&\quad s \in [0,L_\lambda],
\end{split}
\end{equation}
where the tapering function is $E(s)=1-\mathrm{e}^{-k_E^2 s^2}$ and where $k_E$ is a constant that determines how quickly the helix grows to its maximum amplitude $R$ and $k$ is the wavenumber. We use $k_E=k$ in this work. Notably, while many   previous studies have used a right-handed helix and rotated the helix in a  CW direction\cite{higdon1979b,phan1987,ishikawa2007,shum2010,giacche2010,kanehl2014}, in  reality the bacterial flagellum is a left-handed helix that rotates CCW when viewed from behind (though one situation is  just the mirror-image symmetric of the   other). 
\begin{figure}
\centering
\includegraphics[width=0.3\textwidth]{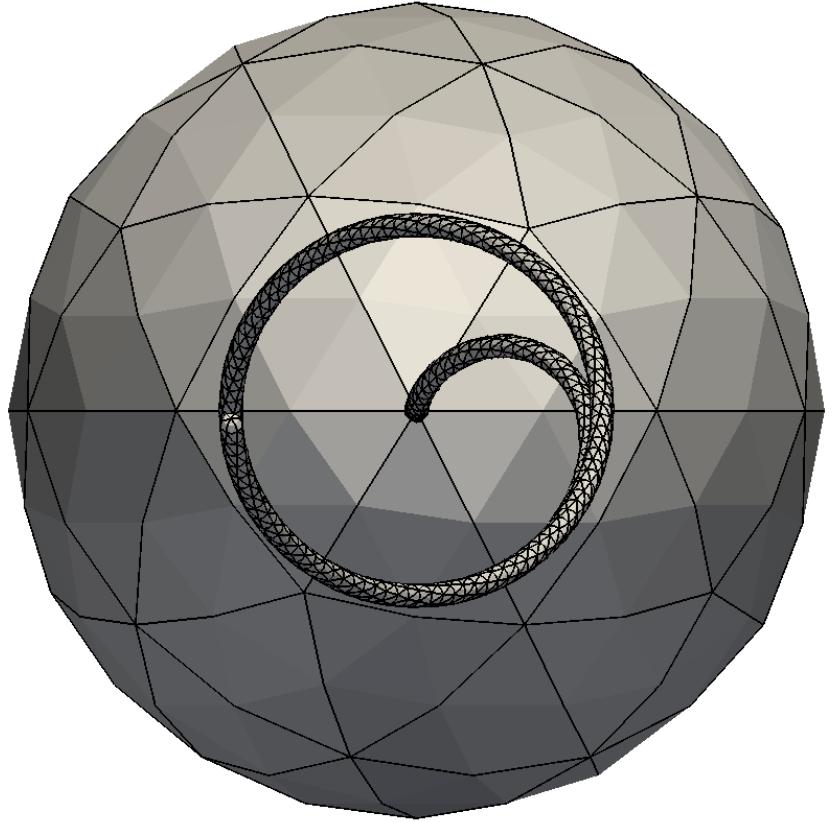}
\caption{Computational model I: Posterior view of the boundary element mesh of an {\it E.~coli} bacterium with the cell body and flagellar filament modelled as a surface distribution of stokeslets on a prolate spheroid and a rigid left-handed helix, respectively. }
\label{fig:backview}
\end{figure}

The axial length of the helix is $L_\lambda$ while its contour length is $L=L_\lambda/\cos(\phi)$, and the angle $\phi= \arctan{(Rk)}$ is the helix pitch angle. Changing the phase angle $\psi$ simply rotates the helix around its axis in the $x$-direction. The linear (or axial) wavelength and wave speed are $\lambda = 2\pi/k$ and $V=\omega/k$ respectively. The curvilinear (or contour) wavelength $\Lambda= \lambda/\cos \phi$ and wave speed $c = V/\cos \phi$ are not constant in the `\textit{end region}' but vary negligibly outside of it. The cross-sectional radius of the helix is $\rho$ and the aspect ratio is defined as $\varepsilon = \rho/L$. The velocity on the surfaces of cell body and flagellum surfaces satisfies
\begin{equation}\label{eq:bemslp-model-1}
\begin{split}
\mathsfbi{u}(\bm{x}_0) = -&\frac{1}{8\pi\mu} \iint\displaylimits_{S_b} \mathsfbi{G}(\bm{x},\bm{x}_0) \bcdot \bm{f}_h (\bm{x}) \, \mathrm{d} S(\bm{x}) \\
-& \frac{1}{8\pi\mu} \iint\displaylimits_{S_f} \mathsfbi{G}(\bm{x},\bm{x}_0) \bcdot \bm{f}_h (\bm{x}) \, \mathrm{d} S(\bm{x}), \\
&  \qquad \qquad \qquad \qquad \qquad \bm{x}_0 \in S_{b,f}.
\end{split}
\end{equation}
If the bacterium is swimming or held in place by an optical trap in an infinite fluid medium, the free-space Green's function $\mathsfbi{G}$ given by equation \eqref{eq:oseenmathsfbi} is used. However, if the bacterium is swimming close to a glass surface or stuck on it, the Oseen-Burgers mathsfbi $\mathsfbi{G}$ is replaced with appropriate wall-modified Green's function $\mathsfbi{G}^w$  satisfying the no-slip velocity condition on the semi-infinite wall \cite{blake1974}. The expression for $\mathsfbi{G}^w$ is provided in appendix~\ref{sec:blakelets}. 

For a free-moving  bacterium, the total hydrodynamic force and torque acting on it should be zero, i.e.
\begin{equation}\label{eq:force-torque-free}
\bm{F}_b + \bm{F}_f = \bm{0}, \quad \bm{T}_b + \bm{T}_f = \bm{0}.
\end{equation}
The torques are calculated about the point where the flagellum is attached to the cell body. The kinematic conditions stipulate that the flagellum remains attached to the  body and thus
\begin{align}
& \bm{u}(\bm{x}_0) = \bm{U} + \bm{\Omega}_b \times \bm{x}_0, \quad \bm{x}_0 \in S_b, \label{eq:kinematic-model-1a} \\
& \bm{u}(\bm{x}_0) = \bm{U} + (\bm{\Omega}_b + \bm{\Omega}_m)\times \bm{x}_0, \quad \bm{x}_0 \in S_f, \label{eq:kinematic-model-1b}
\end{align}
where $\bm{U}$ and $\bm{\Omega}_b$ are the translational and angular velocities of the cell body while $\bm{\Omega}_m$ is the angular velocity of the motor (equal to the relative rotational velocity of the flagellum with respect to the cell body). The lab frame angular velocity of the flagellum is $\bm{\Omega}_f = \bm{\Omega}_b + \bm{\Omega}_m$.  {The angular velocities and torque acting on the cell body and flagellum are measured at the point where the motor is located, $\bm{x}_m$, which is taken to be the origin in our simulations.}

The cell body and flagellum are discretised into $N_b$ and $N_f$ elements respectively. The total number of unknowns to be solved for is $3N_b + 3N_f + 6$, namely the $3N_b+ 3N_f$ components of surface tractions on the cell body and the flagellum and 6 components of the translational and rotational velocity of the cell body. The relative rotational velocity of the flagellum $\bm{\Omega}_m$ serves as the forcing for the system. Equations \eqref{eq:bemslp-model-1} provide us with $3N_b + 3N_f$ equations while equations \eqref{eq:force-torque-free} provide us additional 6 equations.

In the situation where the body of the bacterium is stuck to the wall, we now have $\bm{U}=\bm{0}$ and $\bm{\Omega}_b=\bm{0}$. Therefore, the kinematic conditions simplify to
\begin{align}
& \bm{u}(\bm{x}_0) = \bm{0}, \quad \bm{x}_0 \in S_b, \label{eq:kinematic-model-1a-stuck} \\
& \bm{u}(\bm{x}_0) = \bm{\Omega}_m \times \bm{x}_0, \quad \bm{x}_0 \in S_f. \label{eq:kinematic-model-1b-stuck}
\end{align}
There is a net force and torque required to hold the bacteria stationary. The hydrodynamic traction acting on the cell body and flagellum is unknown. Hence, we need to solve $3N_b + 3N_f$ unknowns using $3N_b + 3N_f$ equations provided by the boundary integral equation \eqref{eq:bemslp-model-1}.

\subsection{Computational Model II: BEM-SBT}\label{sec:cm-II}
\begin{figure}
\centering
\includegraphics[width=0.5\textwidth]{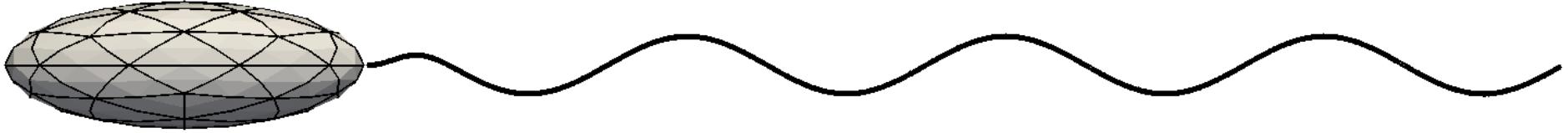}
\caption{Computational model II: Side view of the boundary element mesh of an {\it E.~coli} bacterium with the cell body modelled as a surface distribution of stokeslets on a prolate spheroid and the flagellar filament modelled as a line distribution of stokeslet on the centreline of a rigid left-handed helix.}
\label{fig:sideview}
\end{figure}

In our second computational model (termed CM-II), the cell surface is discretised in the same way as in \S \ref{sec:cm-I} but the flagellar hydrodynamics are described using slender body theory  \cite{johnson1980}, as illustrated in  Fig.~\ref{fig:sideview}. Since the bacterial flagellar filament is a very slender helix of aspect ratio $\varepsilon \sim 0.002$, we can take advantage of past classical work on the dynamics of slender filaments in viscous fluids. Using the formulation of Johnson \cite{johnson1980},  the velocity $\bm{u}$ of the centreline $C_f$ of a slender body is linearly related to the hydrodynamic force per unit length, $\bm{f}_h$, acting on it through the integral relationship,
\begin{equation}\label{eq:SBTequation}
\begin{split}
\bm{u}(\bm{x}_0) = -\frac{1}{8\pi \mu}\bm{\Lambda}[\bm{f}_h](\bm{x}_0)- \frac{1}{8\pi \mu}\bm{K}&[\bm{f}_h](\bm{x}_0), \\
& \bm{x}_0 \in C_f,
\end{split}
\end{equation}
where $ 0 \leq s \leq L$ is the arclength along the centreline of the filament.
 In Eq.~\eqref{eq:SBTequation}, the  local, $\bm{\Lambda}$, and non-local, $\bm{K}$, operators are given by  \cite{johnson1980}
\begin{equation}\label{operators1} 
\begin{split}
\bm{\Lambda}[\bm{f}_h](\bm{x}_0) &= [-c(\mathsfbi{I} + \bm{\hat{s}}\bm{\hat{s}}) + 2(\mathsfbi{I} - \bm{\hat{s}}\bm{\hat{s}})] \bcdot \bm{f}_h(\bm{x}_0),
\end{split}
\end{equation}
\begin{equation}\label{operators2}
\begin{split}
\bm{K}[\bm{f}_h](\bm{x}_0) = \int\displaylimits_{0}^{L} & \left(\bm{f}_h(\bm{x})\bcdot \mathsfbi{G}(\bm{x},\bm{x}_0) \right. \\
& \left. - \frac{\mathsfbi{I} + \bm{\hat{s}}\bm{\hat{s}}}{|s-s^\prime|}\bcdot \bm{f}_h(\bm{x}_0)\right) \,\mathrm{d}s^\prime(\bm{x}),
\end{split}
\end{equation}
where $c=\log{(\varepsilon^2 e)}$. Note that in this formulation, the  cross-sectional radius of the  body varies slowly as $r(s) = 2\varepsilon \sqrt{s(L-s)}$ where $\varepsilon = r(L/2)/L$, ensuring algebraically-accurate results. The cross-sectional radius at the midpoint $s=L/2$ is taken to be equal to the radius of the flagellum, i.e.~$r(L/2) =\rho$ . Equation \eqref{operators2} becomes formally singular when $s=s^\prime$ and this singularity is removed by regularising the integral \cite{tornberg2004}.

The hydrodynamics of the bacterium is then described with a surface and a line distribution of stokeslet, representing the cell body and the flagellum respectively,
\begin{equation}\label{eq:bemslp-model-2a}
\begin{split}
\mathsfbi{u}(\bm{x}_0) = -&\frac{1}{8\pi\mu} \iint\displaylimits_{S_b} \mathsfbi{G}(\bm{x},\bm{x}_0) \bcdot \bm{f}_h (\bm{x}) \, \mathrm{d} S(\bm{x}) \\
-& \frac{1}{8\pi\mu} \int\displaylimits_{C_f} \mathsfbi{G}(\bm{x},\bm{x}_0) \bcdot \bm{f}_h (\bm{x}) \, \mathrm{d} s(\bm{x}), \quad \bm{x}_0 \in S_b,
\end{split}
\end{equation}
and
\begin{equation}\label{eq:bemslp-model-2b}
\begin{split}
\mathsfbi{u}(\bm{x}_0) = -&\frac{1}{8\pi\mu} \iint\displaylimits_{S_b} \mathsfbi{G}(\bm{x},\bm{x}_0) \bcdot \bm{f}_h (\bm{x}) \, \mathrm{d} S(\bm{x}) \\
-&\frac{1}{8\pi \mu}\bm{\Lambda}[\bm{f}_h](\bm{x}_0)- \frac{1}{8\pi \mu}\bm{K}[\bm{f}_h](\bm{x}_0), \\ 
& \quad \qquad \qquad \qquad \qquad \qquad \bm{x}_0 \in C_f.
\end{split}
\end{equation} 

For a free-swimming cell, the  force and torque balance equations are the same as in Eq.~\eqref{eq:force-torque-free}, except that the force and torque on the flagellum are now computed by integrating $\bm{f}_h$ along the centreline,
\begin{equation} \label{eq:force-torque-flagellum}
\begin{split}
\bm{F}_f &=\int\displaylimits_{C_f} \bm{f}_h(\bm{x}) \,\mathrm{d}s(\bm{x}), \\ 
\bm{T}_f &=\int\displaylimits_{C_f} \bm{x} \times \bm{f}_h(\bm{x}) \,\mathrm{d}s(\bm{x}),  \quad \bm{x} \in C_f.
\end{split}
\end{equation}
The velocity of the cell body is same as before, i.e.~Eq.~\eqref{eq:kinematic-model-1a}, while the velocity of the flagellum centreline is now given by,
\begin{equation} \label{eq:kinematic-model-2}
\begin{split}
\bm{u}(\bm{x}_0) = \bm{U} + (\bm{\Omega}_b + \bm{\Omega}_m)\times \bm{x}_0, \quad \bm{x}_0 \in C_f.
\end{split}
\end{equation}
The cell body surface and flagellum centreline are discretised into $N_b$ and $\tilde{N}_f$ elements that requires us to solve for $3N_b + 3\tilde{N}_f + 6$ unknowns.

In the situation where the cell is stuck to the wall, we now have $\bm{U}=\bm{0}$ and $\bm{\Omega}_b=\bm{0}$. The kinematic conditions for the body is same as equation \eqref{eq:kinematic-model-1a-stuck} and that for the flagellum simplifies to,
\begin{equation}
\bm{u}(\bm{x}_0) = \bm{\Omega}_m \times \bm{x}_0, \quad \bm{x}_0 \in C_f. \label{eq:kinematic-model-2-stuck}
\end{equation}
We need to solve $3N_b + 3\tilde{N}_f$ unknowns tractions on the cell body and flagellum using $3N_b + 3\tilde{N}_f$ equations provided by the boundary integral equations \eqref{eq:bemslp-model-2a} and \eqref{eq:bemslp-model-2b}. 

\subsection{Validation of the computational models}\label{sec:validation}
\begin{figure}
\centering
\includegraphics[width=0.49\textwidth]{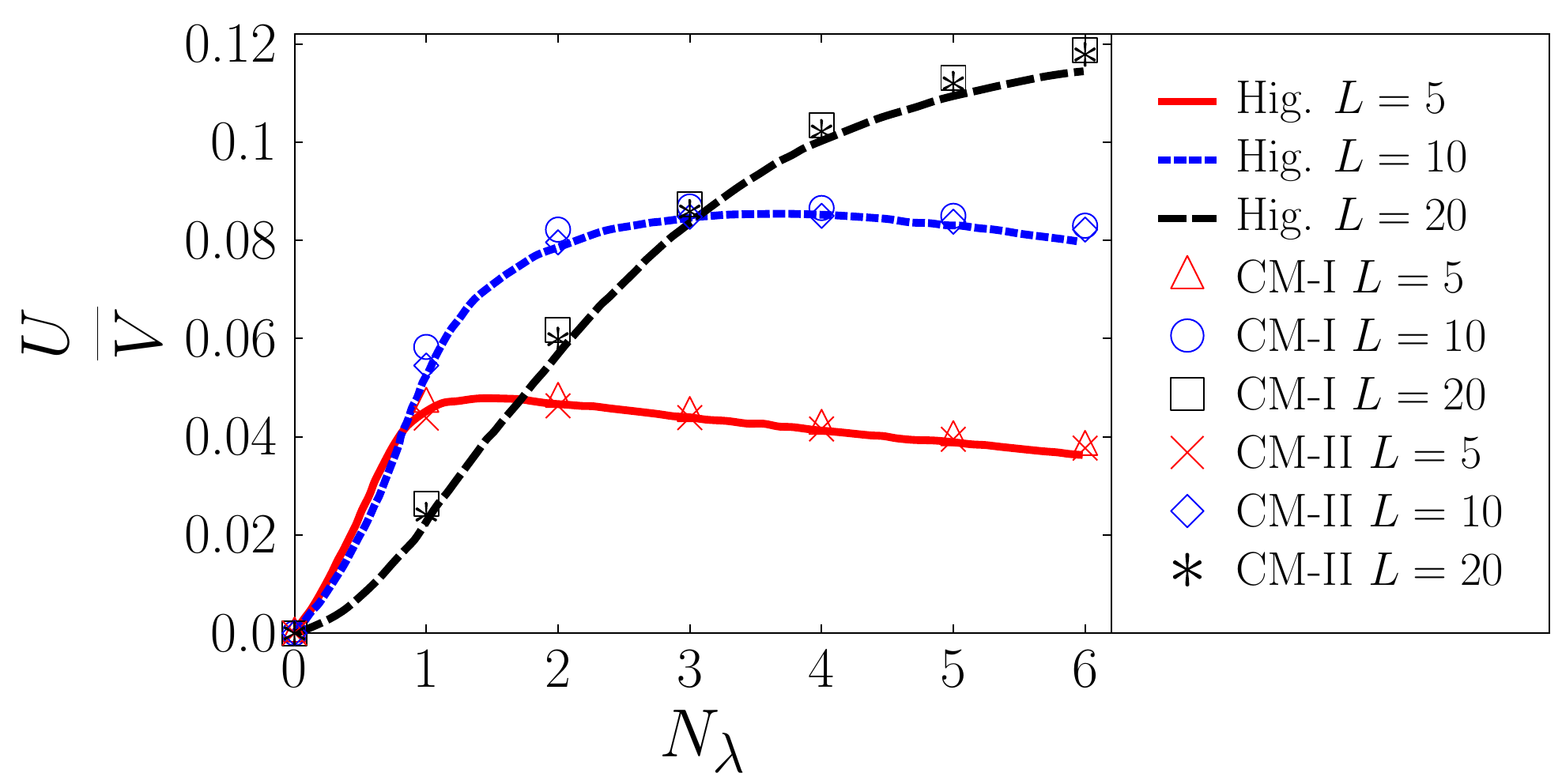}
\caption{Validation of the computational models, CM-I and CM-II, with the semi-analytical results of Higdon \cite{higdon1979b}. Average swimming speed plotted as a function of number of waves with $\rho/a = 0.02$, $\alpha k =1$ and $k/k_E=1$ for three different flagella lengths $L= 5$ (solid red lines), 10 (dotted blue line) and 20 (dashed black line).}
\label{fig:higdon}
\end{figure}

We first validate computational model I by performing simulations relevant for the hydrodynamics of (i) a pair of spheres approaching each other, as detailed in appendix~\ref{sec:spheres} and (ii) a slender cylinder rotating in an infinite fluid and translating next to a wall,  as shown in appendix~\ref{sec:cylinder}. In both case we compare our   numerical results with exact solutions with excellent agreement. 

We next validate both
 computational models using past semi-analytical results for flagellar swimming. In a landmark paper, Higdon \cite{higdon1979b} used slender body theory to model a microorganism swimming by propagating helical waves. Instead of discretising the surface of the cell body, he used modified Green's functions satisfying the no-slip boundary condition on the cell body modelled as a sphere. The swimming speed, $U/V$, normalised by the linear wave speed, is plotted as a function of the number of waves, $N_\lambda$, for three different flagellar lengths with the parameters $\alpha k$ and $k/k_E$ kept fixed in Fig.~\ref{fig:higdon}. Our computational  results are in excellent agreement with the results of Higdon \cite{higdon1979b}.  
 
However, it is important to note a few differences between our two models. First, and unsurprisingly, solving the integral equation on a line (CM-II) instead on a surface (CM-I) results in faster computations.  Second, the surface mesh of a helical flagellum used in computational model I has a circular cross-section of constant radius $\rho$, see appendix~\ref{sec:discretisation}, while the slender body theory used in computational model II a circular cross-section of radius $\rho$ at the centre that slowly tapers towards the ends. Third, in situations where the flagellum is very close to another body, for example a semi-infinite plane wall as in \S\ref{sec:stuck}, we expect near-field hydrodynamic interactions with the wall to play an important role.
In these situations, slender body theory will fail to resolve the  hydrodynamics, necessitating full surface discretisation. This makes computational model I a more accurate but also a costlier approach. For example, simulations of swimming bacteria, presented in \S \ref{sec:swimming} using CM-I, were performed with $N_b=80$ and $N_f = 4620$ leading to a system size of 14106, requiring $\sim$ 74s to be solved. On the other hand, simulations using CM-II were performed with $N_b=80$ and $\tilde{N}_f = 300$ leading to a system size of 1146, requiring $\sim$ 1s to be solved. These simulations were performed on a computer with 3401 MHz CPU clock speed and 16GB RAM. 

 %%%%%%%%%%%%%%%%%%%%%%%%%%%%
\section{Results}\label{sec:results}
\begin{table}[t]
{\small
  \begin{tabular*}{0.48\textwidth}{@{\extracolsep{\fill}}lllll}  
    \hline
	Data & Body Length & Body Width & Axial Length & Motor Speed \\
	set  & (2a) & (2b) & ($L_\lambda$) &  ($\Omega_m$)  \\
	\hline
    A & \SI{2.5}{\micro\metre} & \SI{0.88}{\micro\metre} & \SI{8.3}{\micro\metre} & $2\pi \times 154 s^{-1}$ \\
    B & \SI{2.0}{\micro\metre} & \SI{0.86}{\micro\metre} & \SI{10.0}{\micro\metre} & $2\pi \times 87 s^{-1}$ \\
	C & \SI{2.5}{\micro\metre} & \SI{0.88}{\micro\metre} & \SI{6.2}{\micro\metre} & $2\pi \times 111 s^{-1}$  \\
    \hline
  \end{tabular*}
\label{tbl:data}}
\caption{Geometrical and kinematic characteristics of the three data sets  from   Darnton et al.~\cite{darnton2007}. Data sets A and B are measurements of  swimming \textit{E.~coli} cells in two different fluids using a bundle of flagellar filaments, while C is that of a stuck 
    \textit{E.~coli} bacterium with a single rotating flagellum. The length and angular velocity scales for the problem are $a$ and $\Omega_m$, respectively}
\end{table}

We use our computational models to address three sets of data for \textit{E. coli} from the experiments of Darnton et al.~\cite{darnton2007}. Data sets A and B are concerned with \textit{E. coli} bacteria swimming using a  bundle  of helical flagellar filaments in two different fluids and our numerical results in this case are given in \S \ref{sec:swimming}. In contrast, the bacteria in data set C are stuck on a glass surface with only one rotating flagellum, with our results provided in \S \ref{sec:stuck}. While  the main goal of our paper is to use the experiments on stuck bacteria to infer the value of the motor torque, we also model  the case of swimming cells to compare our simulations with observable physical quantities in the experiments, namely the cell swimming speed and body rotation frequency.

In all cases, the helical flagellar filaments have identical pitch, $\lambda=$ \SI{2.22}{\micro\metre}, helical radius,  $R=$ \SI{0.2}{\micro\metre} and cross-sectional radius, $\rho=$ \SI{0.012}{\micro\metre}. The other relevant parameters including geometrical dimensions and motor rotation speed   vary from one data set to the next and are listed in Table \ref{tbl:data}.  The number of filaments in the bundle for data sets A and B is unknown and we focus here on the case of a single filament (see discussion below on the role played by the filament radius). In order to model a bacterium, one needs to transform the sphere in \S \ref{sec:validation} into a prolate spheroid of  appropriate dimensions. The motor rotation speed, $\Omega_m$, and the major semi-axes length, $a$, are used as the angular velocity and length scale, respectively (see Table \ref{tbl:data}). The major axes of the cell body and the flagella are aligned with each other along the $x$-axis. We maintain a small gap between the cell body and the helix $\sim $ \SI{0.0125}{\micro\metre}, thereby avoiding a singularity in the boundary integral equation (results are unaffected by alterations in this distance). Each calculation presented below has been averaged over the phase angle $\psi$. 
 
 Numerically, all  results in the following sections are computed with $N_b=80$, $N_f = 4620$ and $\tilde{N}_f = 300$. We have performed convergence tests and verified that finer grid resolution than these values produce negligible changes in our results. 
 
\subsection{Swimming bacteria} \label{sec:swimming}

Using   computational models I and II and first assuming that the cell is propelled by a single helical filament, we prescribe the dimensionless motor rotation velocity along the major axis of the cell body, $\bm{\Omega}_m=\bm{\hat{i}}$ that acts as the forcing on the system (see Table \ref{tbl:data} for the dimensional values). The observable physical quantities   are the swimming speed, $\bm{U}$, and the angular velocity of the cell body, $\bm{\Omega}_b$, occuring in the negative $x-$direction since the body counter-rotates compared to the flagella. Data set A takes place in the  fluid medium termed motility buffer   (MB+) for which the viscosity is 0.93 cP. The swimming speed $U$ computed using the two computational models I and II, \SI{20.6}{\micro\metre}$/s$ is lower than that measured in experiments, \SI{29}{\micro\metre}$/s$. The angular velocity of the body $\Omega_b$ using models I and II are 27 Hz and 25 Hz respectively, slightly higher than that measured in experiments, 23 Hz. The hydrodynamic torque experienced by the flagella, equal to the motor torque,  were found to be 728 and 682 pN~nm with models I and II, respectively.

 Data set B takes place in the  MB+ fluid  with added methylcellulose (MC) for which the viscosity is 3.07 cP. The cell body length and motor rotation speed are smaller while the flagella is longer in this case when compared to data set A. The swimming speed $U$ computed using the two computational models I and II,  is found to be \SI{12.6}{\micro\metre}$/s$, significantly lower than that measured in experiments, \SI{31}{\micro\metre}$/s$. The angular velocity of the body $\Omega^b$ using models I and II are 21.3 Hz and 20.1 Hz respectively and agree very well with that measured in experiments equal to 21 Hz. The hydrodynamic torque experienced by the flagella in this case were found to be 1504 and 1418 pN~nm with models I and II, respectively. These results are summarised in Table \ref{tbl:swimmingexpt}. 
 
\begin{table}[t]
{\small
  \begin{tabular*}{0.48\textwidth}{@{\extracolsep{\fill}}llll}  
    \hline
	Data set A & Expt & CM-I & CM-II \\
  	\hline
    Body rotation speed (Hz) & 23 & 27 & 25 \\
    Filament rotation speed (Hz) & 131 & 127 & 129 \\
	Cell speed (\SI{}{\micro\metre}$/s$) & 29 & 20.6 & 20.6 \\
	Motor torque (pN~nm) & - & 728 & 682 \\
    \hline
	Data set B  & Expt & CM-I & CM-II \\
    \hline
    Body rotation speed (Hz) & 21 & 21.3 & 20.1 \\
    Filament rotation speed (Hz) & 66 & 65.7 & 66.9 \\
	Cell speed (\SI{}{\micro\metre}$/s$) & 31 & 12.6 & 12.6 \\
	Motor torque (pN~nm) & - & 1504 & 1418 \\
	\hline
  \end{tabular*}
\label{tbl:swimmingexpt}}
    \caption{Comparison between measurements and   computations for data sets A and B relevant for swimming bacterium in fluids MB+ and MB+ with MC, respectively, from  Darnton et al.~(2007) \cite{darnton2007}. Data set A has dimensionless body length $2a=2$, width $2b=0.7$, filament axial length $L_\lambda=8.3$ and viscosity = $0.93$ cP.  Data set B has dimensionless body length $2a=2$, width $2b=0.86$, filament axial length $L_\lambda=10.0$ and viscosity = $3.07$ cP. The motor rotation speed is prescribed in the simulations}
\end{table}

\begin{figure}
\centering
\includegraphics[width=0.43\textwidth]{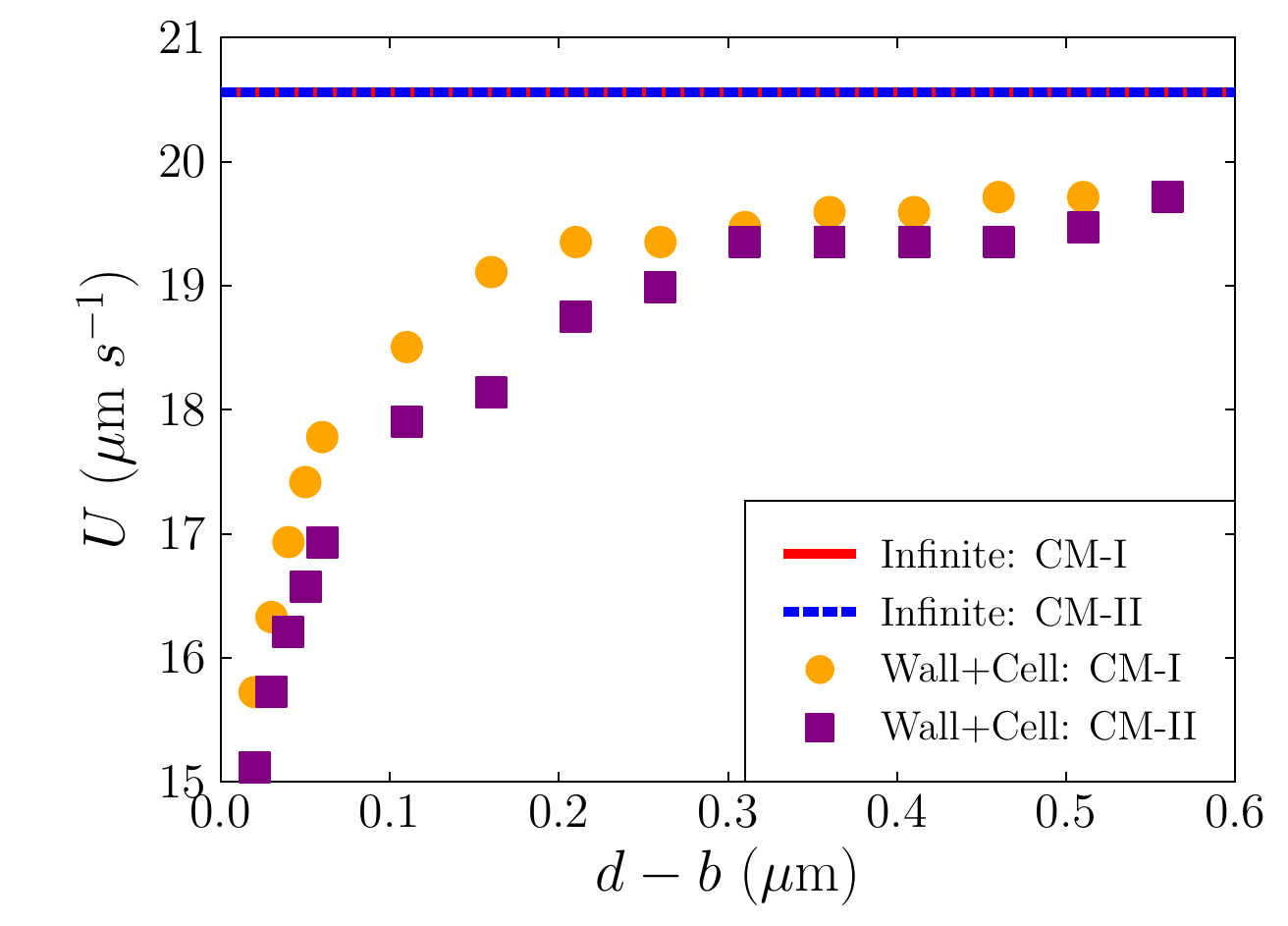}
\includegraphics[width=0.43\textwidth]{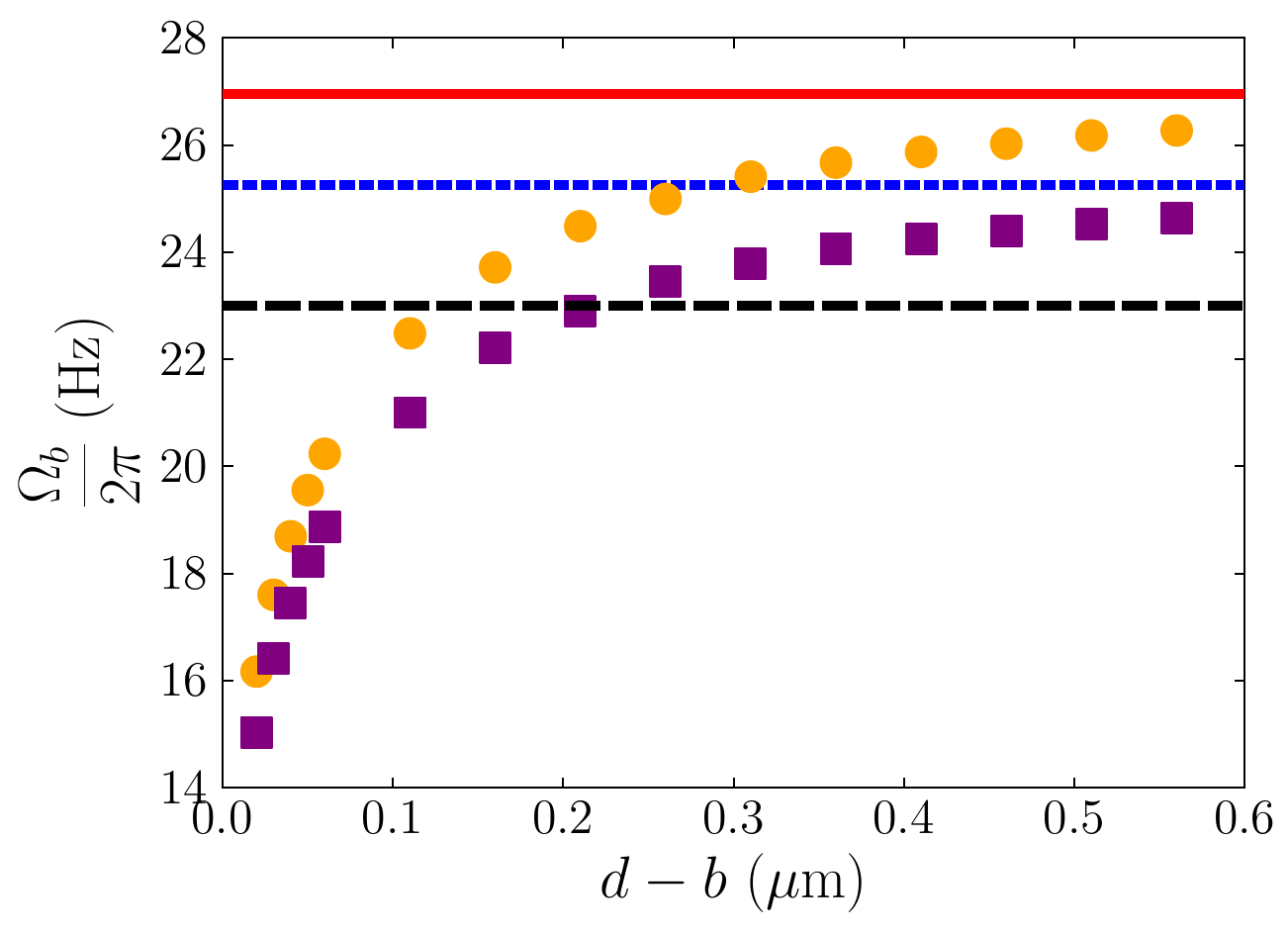}
\includegraphics[width=0.43\textwidth]{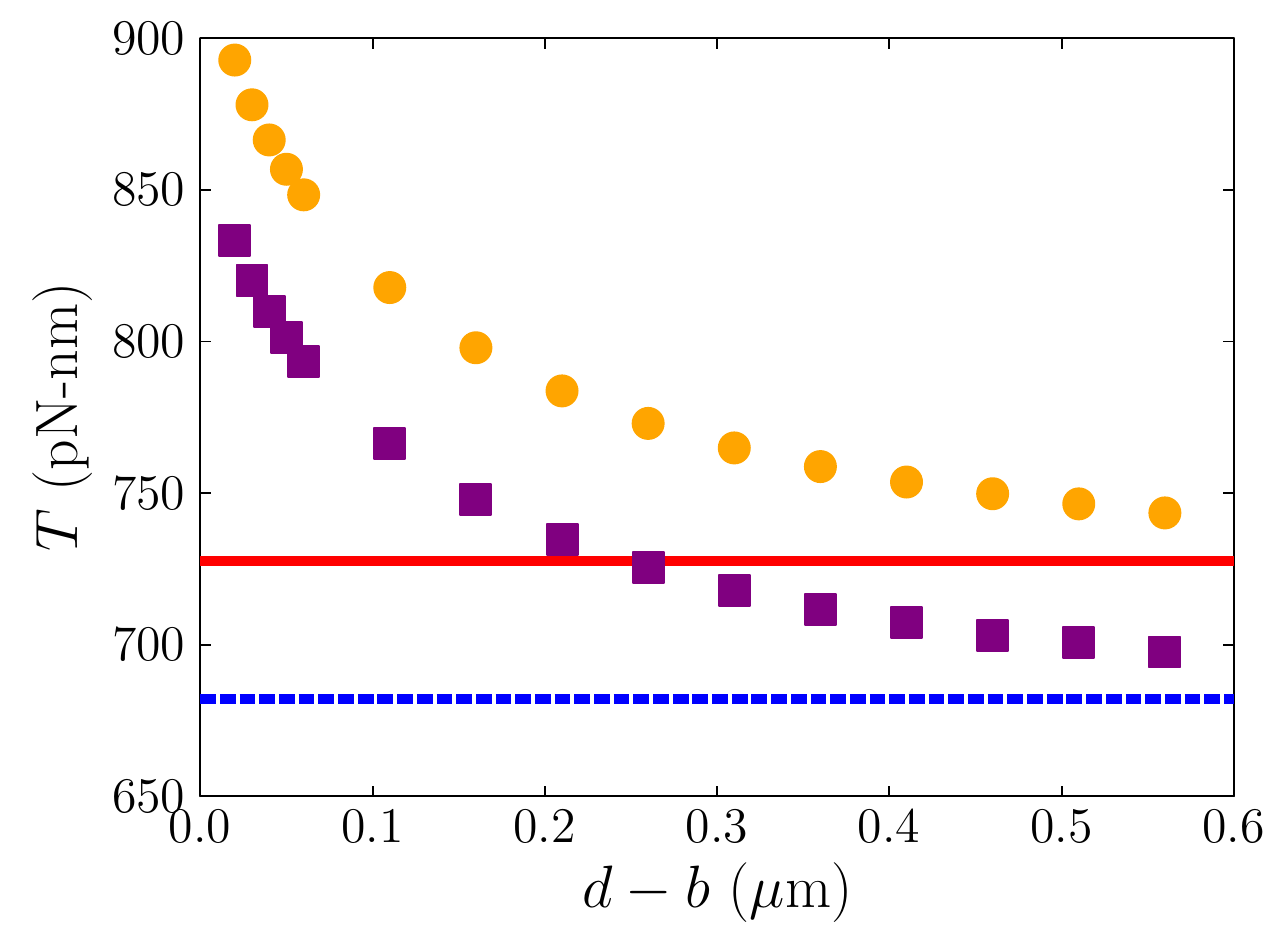}
\caption{Top: Variation of the bacterium swimming speed, $U$ (\SI{}{\micro\metre}$/s$),  with the minimum distance between the cell body and the wall, $d-b$ (\SI{}{\micro\metre}), for data set A. Circles (wall effect) and solid red line, \SI{20.6}{\micro\metre}$/s$ (infinite medium) correspond to model I while squares (wall effect) and dotted blue line, \SI{20.6}{\micro\metre}$/s$ (infinite medium) correspond to model II.  The swimming speed in the  experiments of Darnton et al.~\cite{darnton2007} was found to be \SI{29}{\micro\metre}$/s$ (not shown). 
Middle: Dependence of the  rotational speed of the cell body,  $\Omega_b/2\pi$ (Hz), with the distance to the wall. Circles (wall effect) and solid red line, 27~Hz (infinite medium) correspond to model I while squares (wall effect) and dotted blue line, 25~Hz (infinite medium) correspond to model II. Black dashed line corresponds to rotational speed of 23~Hz measured in experiments of Darnton et al.~\cite{darnton2007}.
Bottom: Variation of the motor torque, $T$ (pN~nm), with the minimum distance between the cell body and the wall. Circles (wall effect) and solid red line, 728~pN~nm (infinite medium) correspond to model I while squares (wall effect) and dotted blue line, 682~pN~nm (infinite medium) correspond to model II.
}
\label{fig:swimmingwall}
\end{figure}

It is not specified in the experiments for data sets A and B whether the bacteria are swimming close to a surface. In order to investigate  how the presence of a wall may effect the swimming kinematics, we vary the distance of the bacterium from the wall while keeping the rotation rate of the motor  fixed as in the infinite fluid medium case. The free-space Green's function $\mathsfbi{G}$ used in the models is replaced with $\mathsfbi{G}^w$ to account for the no-slip condition on the wall. The variation of the swimming speed, rotational speed and motor torque with the minimum distance between the cell body and wall $d-b$ (see notation in Fig.~\ref{fig:schematic}), are shown in Fig.~\ref{fig:swimmingwall} for data set A (the same trends are seen for data set B, the corresponding plots of which are not presented here for brevity). 

As expected, we find that the swimming and rotational speed of the cell body decrease monotonically while the motor torque increases as the cell is moved closer to the wall. As the distance between the bacterium and the wall increases, the computed velocities and torque asymptotically reach their infinite fluid medium values. 

In the freely swimming cells experiments of Darnton et al.~\cite{darnton2007}, it is observed that some flagella can wrap around the cell body. While the distance between a wrapped flagellum and the cell body is unknown, we may estimate its effect on the torque generation by assuming a worst-case scenario where this distance is  \SI{0.1}{\nano\metre}. The viscous torque arising from lubrication forces due to rotation of the flagellum near a cell body can be estimated using Jeffrey's analytical result for the torque per unit length, $T_h$, experienced by a cylinder of radius $\rho$ rotating with  angular velocity $\Omega$  at a distance $d$ from a plane wall, namely $T_h = -4 \pi \mu \Omega \rho^2 d/ \sqrt{d^2 - \rho^2}$\cite{jeffrey1981}. Assuming that the length of the flagellum rotating next to the cell body is approximately half the cell body length, $a$, and substituting in this formula the relevant values from Tables \ref{tbl:data} and \ref{tbl:swimmingexpt}, we obtain additional torques for data set A and B of 16 pN~nm and 7 pN~nm respectively. These values do not add significantly to the torque values obtained from the simulations.

Quite remarkably, and except for the value of the swimming speed  in data set B, the agreement between our  simulations with a single flagellar filament and the experiments is excellent. Since the number of filaments in the  helical bundle  is unknown in the experiments, we then use an alternate method to account for the effect of  multiple filaments. In a controlled experimental study\cite{kim2004}, it was shown that the flow field generated by a bundle of two filaments is well approximated by the flow generated by a single rigid helix with twice the filament radius. 
  For data sets A and B, we thus used computational model I to carry out additional simulations  varying the cross-sectional radius of the helical  filament, taking values of either $2\rho$, $3\rho$ or $4\rho$, where $\rho =$ \SI{0.012}{\micro\metre} is the radius of an isolated filament. The results of these simulations are summarised in Table~\ref{tbl:swimmingexptthick}. We find that, as the thickness of the filament increases, the mismatch between our numerical results and the experiments increases, indicating that the hydrodynamics of a bundle is closer to that of an isolated filament. 

\begin{table}[t]
{\small
  \begin{tabular*}{0.48\textwidth}{@{\extracolsep{\fill}}lllll}  
    \hline
	Data set A & Expt & $2\rho$ & $3\rho$ & $4\rho$ \\
  	\hline
    Body rotation speed (Hz) & 23 & 32.5 & 37.4 & 42.2 \\
    Filament rotation speed (Hz) & 131 & 121.5 & 116.6 & 111.8\\
	Cell speed (\SI{}{\micro\metre}$/s$) & 29 & 20.6 & 19.4 & 18.1 \\
	Total   torque (pN~nm) & - & 877 & 1012 & 1144 \\
    \hline
	Data set B  & Expt & $2\rho$ & $3\rho$ & $4\rho$ \\
    \hline
    Body rotation speed (Hz) & 21 & 25.3 &  28.7 & 31.9 \\
    Filament rotation speed (Hz) & 66 & 61.7 & 58.3 & 55.1 \\
	Cell speed (\SI{}{\micro\metre}$/s$) & 31 & 12.0 & 10.9 & 10.4 \\
	Total   torque (pN~nm) & - & 1784 & 2026 & 2256 \\
	\hline
  \end{tabular*}
\label{tbl:swimmingexptthick}}
    \caption{Comparison between measurements   and computations using CM-I for data sets A and B relevant for swimming bacterium in fluids MB+ and MB+ with MC, respectively, from  Darnton et al.~(2007) \cite{darnton2007} with varying cross-sectional radius of the effective helical filament (from twice the radius of a single filament to four times); all other parameters as in Table \ref{tbl:swimmingexpt}}
\end{table}

The discrepancy between the experimental and our computational results with a single filament may be attributed to the fine details of the filament interactions. Firstly, the effect of hydrodynamic interactions between the  filaments  is absent in the simulations. Secondly, steric interactions between filaments could also play a significant role \cite{hu2015}. However, it was noted previously that  including filament interactions does not lead to a linear increase in swimming speeds or rotational speeds, both seen in experiments \cite{darnton2007} and numerical simulations \cite{hu2015}. The viscosity for data set B is three times that of data set A but it is  believed that the motor torque is independent of the medium viscosity. Hence, it is quite surprising that the motor torque for data set B is almost twice as high as that for data set A. It might be that the fluid medium surrounding the rotating flagellum is not perfectly Newtonian and the presence of polymers could induce additional stresses in the flow, despite the fact that  the \SI{24}{\nano\metre} diameter flagellar filament might be comparable in size to both the radius of gyration of the polymer molecules and to the  distance between them.

\subsection{Bacteria stuck on a surface}\label{sec:stuck}

The most important result of our study is the situation where the bacterium is stuck on a (glass) surface with an immobile cell body rotating a single flagellar filament, termed data set C in Table \ref{tbl:data}. All other geometric and kinematic parameters are known in this case, apart from the distance to the surface, making it an ideal situation to model in order to predict the value of the torque exerted by a single rotary motor.  
The forcing for this system is $\bm{\Omega}_f=\bm{\Omega}_m$  but unlike the swimming bacterium case there are no direct observables in the experiments to compare with as the cell body is not moving, i.e.~$\bm{U}=\bm{0}$ and $\bm{\Omega}_b=\bm{0}$. Note that the bacterium is not force and torque free in this case.

We first consider a flagellum rotating in an infinite fluid medium in the absence of the cell body. Using computational models I and II, we get motor torque values equal to 474 pN~nm and 438 pN~nm, respectively which are significantly higher than that obtained by resistive force theory calculations, 370 pN~nm \cite{darnton2007} (and reproduced in Table \ref{tbl:torqueexpt}). We then place an immobile cell body that simply acts as a hydrodynamic obstacle to the rotating flagellum and find the torques increase negligibly to 475 pN~nm and 440 pN~nm for models I and II respectively. 
 The increase in the force experienced by the flagella, $\sim 6\%$ is also   small but higher than the increase in the torque, $\sim 0.2\%$. These results prove that an immobile cell body does not produce significant hydrodynamic resistance to the rotating flagellum.  

\begin{figure}
\centering
\includegraphics[width=0.43\textwidth]{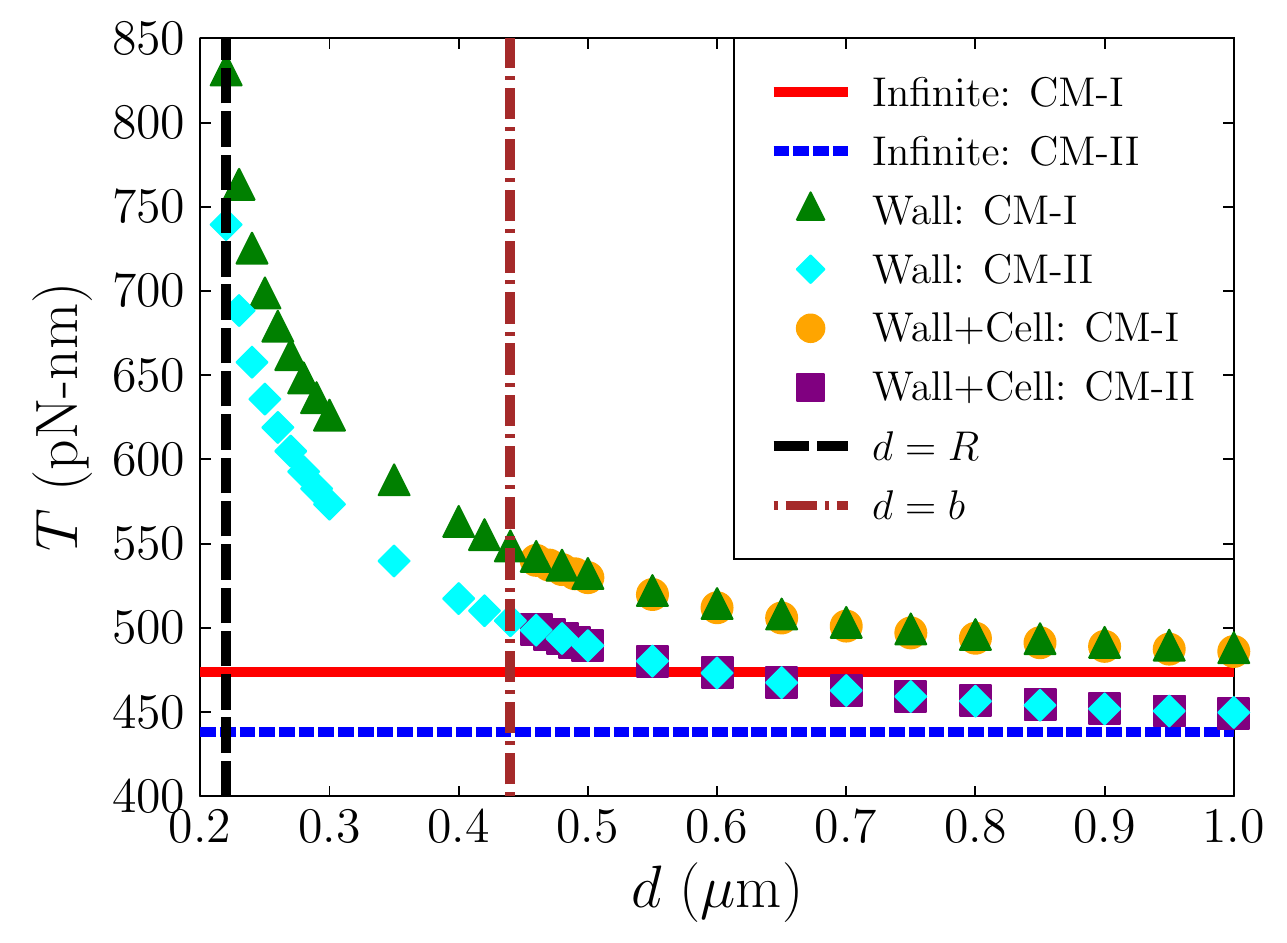}
\includegraphics[width=0.43\textwidth]{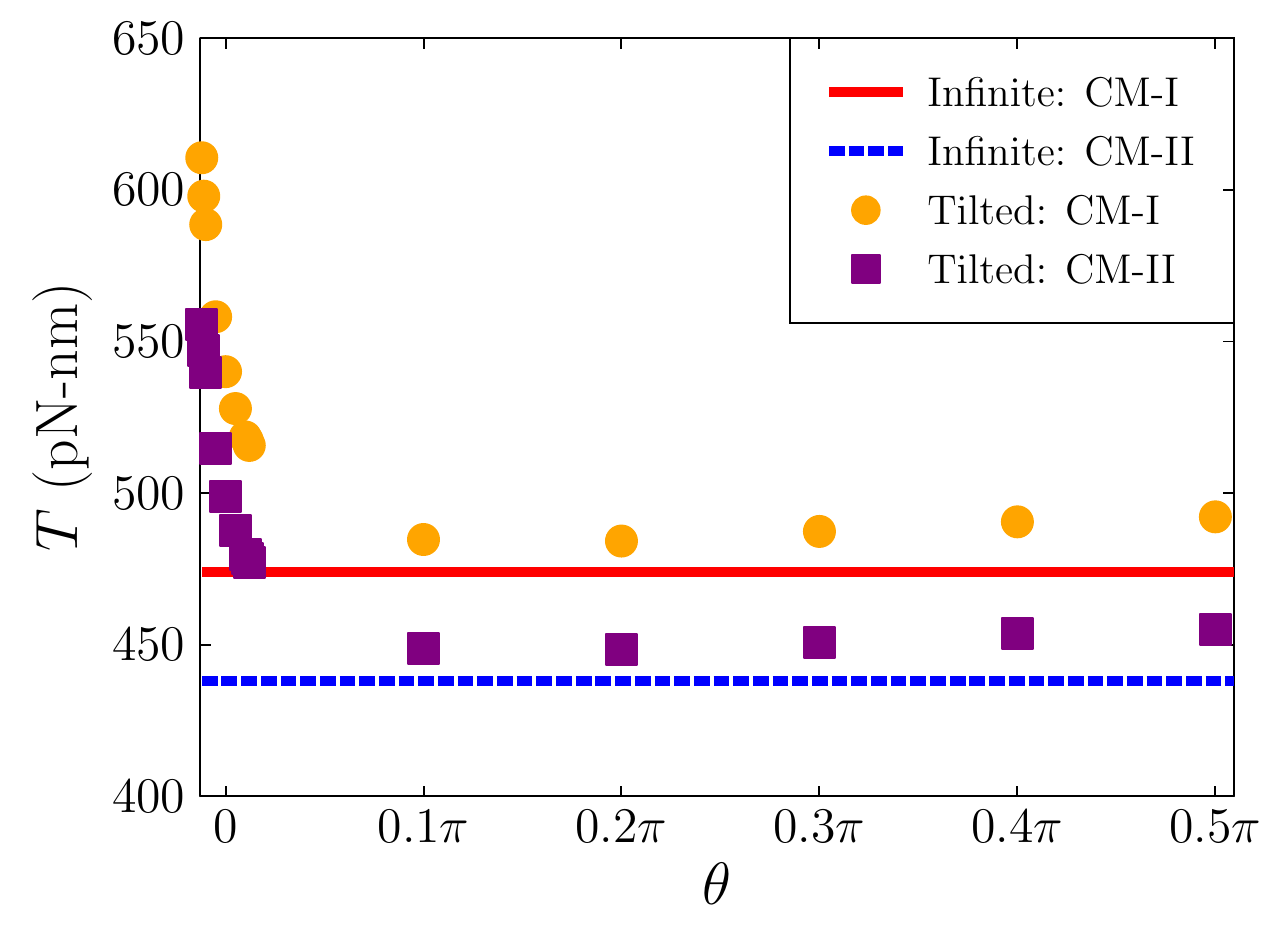}
\caption{Top: Variation of motor torque $T$ (pN~nm) as a function of the distance $d$ (\SI{}{\micro\metre}) of the flagellum's centreline from the wall. Triangles and diamonds correspond to motor torque values in the presence of wall but without a cell body, circles and squares correspond to those in the presence of both wall and cell body while the red and blue lines correspond to torque values in an infinite fluid medium without wall or cell body. The vertical black dashed line is $d=R=$ \SI{0.2}{\micro\metre} while the vertical dash-dotted brown line is at $d=b=$ \SI{0.44}{\micro\metre}. Bottom: Variation of motor torque, $T$ (pN~nm), as a function of the tilt angle, $\theta$, of the flagellar axis relative to  the wall. The distance between the major axis of the cell body  and the wall is kept constant at $d=$ \SI{0.46}{\micro\metre}.}
\label{fig:walleffect}
\end{figure}

We next consider the effect of the wall on the torque experienced by a flagellum with and without the cell body while keeping the flagellum parallel to the wall and the   rotation speed of the motor fixed at the experimentally-measured value of $2\pi \times 111$ Hz. As shown in    Fig.~\ref{fig:walleffect} (top), the value of the motor torque  increases as we get closer to the wall.  We may also place a cell body next to the rotating flagellum, exactly reproducing the experimental conditions of a bacterium stuck on a glass surface. The results of computational model I in Fig.~\ref{fig:walleffect} (top) are plotted using triangles when there is no cell body and circles if we add the cell body; similarly,  for computational model II we use  diamond (cell body absent) and squares (cell body present). For both models, we see the two sets of symbols overlap with  each other, indicating that the torque values is essentially unaffected by the  presence of the cell body, consistent with our results in the infinite fluid medium case. The maximum  values of motor torque obtained for computational models I and II are 829 pN~nm and 739 pN~nm when the flagellum is placed very close to the wall at a distance $d=$ \SI{0.22}{\micro\metre}, slightly above the mathematically minimum gap between the flagellum and the wall of \SI{0.02}{\micro\metre}. 

To give an indication of the magnitude of the torque for a more realistic value of the distance to the wall, we may pick the value $d=\SI{0.46}{\micro\metre}$, so that the gap between the cell body and the wall is \SI{0.02}{\micro\metre}.  
In this case, the value of the  motor torque  with and without  the cell are 540~pN~nm and 539~pN~nm for model I and 498 pN~nm in both cases for model II. Note that in either case we can not have zero distance between the wall and the flagellum or the cell body as the integral equations become singular. The full dependence of $T$ on the value of $d$ is shown in Fig.~\ref{fig:walleffect} (top). 

Just as for freely swimming cells, it is likely that in some cases a portion of the flagellum is wrapped around the cell body. We can carry out an  analysis similar to the one in \S \ref{sec:swimming} using the same assumptions to find that the lubrication torque due to a flagellum rotating close to a cell body is now on the order of 11 pN~nm (a smaller value than above due to a smaller rotation frequency). Here again, this contribution to the torque is small and can be neglected.  

\begin{table}[t]
{\small
  \begin{tabular*}{0.48\textwidth}{@{\extracolsep{\fill}}lll}  
    \hline
	Geometrical parameter (\SI{}{\micro\metre}) & CM-I (pN~nm) & CM-II (pN~nm)\\
  	\hline
    $a=2.5 + 10\%$  & 540 + 0.0$\%$ & 498 + 0.0$\%$ \\
    $a=2.5 - 10\%$  & 540 + 0.0$\%$ & 498 + 0.0$\%$ \\
    \hline
    $b=0.88 + 10\%$  & 540 - 2.0$\%$ & 498 - 2.0$\%$ \\
    $b=0.88 - 10\%$  & 540 + 2.6$\%$ & 498 + 2.6$\%$ \\
    \hline
    $\rho=0.012 + 10\%$  & 540 + 3.1$\%$ & 498 + 2.6$\%$ \\
    $\rho=0.012 - 10\%$  & 540 - 3.2$\%$ & 498 - 2.7$\%$ \\
    \hline
	$\lambda=2.22 + 10\%$  & 540 + 0.8$\%$ & 498 + 0.9$\%$ \\
    $\lambda=2.22 - 10\%$  & 540 - 1.1$\%$ & 498 - 1.2$\%$ \\
	\hline
	$R=0.2 + 10\%$  & 540 + 18.5$\%$ & 498 + 18.8$\%$ \\
    $R=0.2 - 10\%$  & 540 - 17.2$\%$ & 498 - 17.5$\%$ \\
	\hline
	$L_\lambda=6.2 + 10\%$  & 540 + 10.5$\%$ & 498 + 10.5$\%$ \\
    $L_\lambda=6.2 - 10\%$  & 540 - 10.5$\%$ & 498 - 10.5$\%$ \\
    \hline
  \end{tabular*}
\label{tbl:uncertain}}
    \caption{Variation in the value of the motor torque with changes in the geometrical parameters of the stuck bacterium with a rotating flagellum next to a wall for both models. The minimum distance between the cell body and the wall is kept fixed at \SI{0.02}{\micro\metre}}
\end{table}

\subsubsection{Tilted flagellum}

The exact orientations of the flagellar filaments relative to the surface have not been reported in the experiments but they  are likely to play an important role. Keeping the cell body's major axis fixed at a distance of $d=$ \SI{0.46}{\micro\metre} from the wall, we vary in our simulations the value of the tilt angle, $\theta$, between the axis of the flagellar filament and the direction of the surface. The results are shown in Fig.~\ref{fig:walleffect} (bottom).  Note that the minimum value of tilt towards the wall is dictated by the minimum separation distance of the flagellum from the wall; specifically in the simulations, we limit ourselves to $\theta_{\mathrm{min}} = -0.012\pi$ corresponding to a minimum distance of \SI{0.0165}{\micro\metre} of the flagellum from the wall. 
 Unsurprisingly, we find that the axial torque required to rotate the flagellum at a constant angular velocity (111~Hz) systematically increases with an increase of tilt angle towards the wall as parts of the flagellum move closer towards it. The maximum values of the torque  at the minimum tilt angle are equal to 611 pN~nm and 556 pN~nm for models I and II respectively, almost 30\% larger than the values when the filaments are tilted  away and far from the surface. These results demonstrate the importance of the flagellum orientation on the motor torque values.

\subsubsection{Geometrical parameters uncertainties}

In addition to intrinsic biological variability, there are unavoidable uncertainties in the measurements of geometrical parameters of the bacteria.   In order to estimate the impact of these uncertainties on our results, we look at the variability of the motor torque, $T$, induced by a $\pm 10\%$ variability in the geometrical parameters, namely $a$, $b$, $\rho$, $\lambda$, $R$ and $L_\lambda$. All the simulations are performed with the minimum gap between the cell body and the wall kept fixed at \SI{0.02}{\micro\metre} and the results are summarised in Table \ref{tbl:uncertain}. We find that variations in the length and width of the cell body and the cross-sectional radius and pitch of the flagellum have negligible effect on the motor torque value. In contrast, the torque varies linearly with the axial length of the flagellum (consistent with the asymptotic resistive force theory \cite{lauga2009}) and varies the most with changes in radius of the helical filament.

 %%%%%%%%%%%%%%%%%%%%%%%%%%%%
\section{Motor torque: Comparison with past experiments and simulations}\label{sec:comparison}

Our simulations in the case of bacteria stuck on a surface exactly reproduce the experiments of Darnton et al.~\cite{darnton2007} and allow us therfore to provide a computational prediction for the value of the motor torque. Our results, shown in Fig.~\ref{fig:walleffect}, were seen to range between 440~pN~nm (minimum value, in bulk fluid) to 829~pN~nm (maximum value in the extreme case where the flagellum all but touches the surface). In order to compare with past experiments and simulations, we summarise in  Table \ref{tbl:torqueexpt} a list of the previous studies that have attempted to find the value of the motor torque of \textit{E. coli}. 

\begin{table}[t]
{\small
  \begin{tabular*}{0.48\textwidth}{@{\extracolsep{\fill}}lll}
    \hline
     & Motor Torque  & Experiments  \\
 Reference   &(pN~nm)& vs.~Numerics \\
    & & vs.~Theory \\
    \hline
    Berry and Berg (1997) \cite{berry1997} & $\sim$ 4,500 & exp \\
    Fahrner et al.~(2003) \cite{fahrner2003} & 1,370 $\pm$ 50 & exp \\
    Chattopadhyay et al.~(2006) \cite{chattopadhyay2006} & 500& exp  \\ 
    Reid et al.~(2006) \cite{reid2006} & 1,260 $\pm$ 190& exp  \\    
	{Darnton et al.~(2007)} \cite{darnton2007} & {370 $\pm$ 100}& exp + th \\ 
    Shimogonya et al.~(2015) \cite{shimogonya2015} & $\sim$ 700& exp + num  \\
    {Hu et al.~(2015)} \cite{hu2015} &  { $\sim$ 1,200}& num \\  
    {Kong et al.~(2015)} \cite{kong2015} & {$\sim$ 1,600} & num \\
    Van Oene et al.~(2017) \cite{van2017}  & %444 $\pm$ 366 -
     874 $\pm$ 206& exp  \\
	{Present work} & {440--829 }& num \\	
    \hline
  \end{tabular*}
\label{tbl:torqueexpt}}
    \caption{List in  chronological order of past experiments (exp) and numerical simulations (num) or theory (th) investigating the value of the bacterial motor torque with range of reported values}
\end{table} 
 
\subsection{Comparison with past experiments}\label{sec:compareexpts}

The early experimental investigation of bacterial motor torque were done with electrorotation method \cite{berg1993b}, however, absolute values of torque were not reported. Since it is difficult to measure the high rotation rates of individual flagella, Berry and Berg \cite{berry1997} tethered a single flagellum to a glass surface and measured the  slowly rotating cell body of  \textit{E.~coli} cells. The rotation of the cell body was stopped by using optical tweezers thereby providing an estimate of the motor torque $\sim$ 4,500 pN~nm. This high value   appears to be an outlier in the literature.

The preferred method used in experiments since then involves shearing most of the flagellar filament and tethering a small spherical bead to the remaining flagellar stub. The rotation rate of the sphere is then measured and the viscous torque acting on it directly relates to the motor torque. Chen and Berg \cite{chen2000} pioneered this tethered bead experimental method, however, they did not give the absolute torque values but  only  relative value of the torques at different frequencies. While  Chen and Berg \cite{chen2000} used spherical beads of diameter \SI{0.45}{\micro\metre} and changed the load by changing the viscosity of the medium, Fahrner et al.~\cite{fahrner2003} used the same method but used beads of different sizes in order to change the load. Unfortunately, neither the value of  torque nor the viscosity of the fluid medium are mentioned in the article. In the low-speed, high-torque limit with spheres whose diameters ranged from 1.0--\SI{2.1}{\micro\metre}, they obtained rotation rates ranging from 78 to 8.6 Hz. A mean torque of $\sim 1,370$ pN~nm is however mentioned by the same group in Darnton et al.~\cite{darnton2007}. 
Finally, using spherical beads of diameter \SI{1}{\micro\metre}  for the tethered bead method, Reid et al.~\cite{reid2006} found the motor torque to be $\sim 1,260$ pN~nm corresponding to a motor speed of 63 $\pm$ 7 Hz. 

A different method was proposed
Chattopadhyay et al.~\cite{chattopadhyay2006} who used optical traps to prevent  \textit{E.~coli} bacteria from swimming. The optical trap force compares directly with the thrust force that propels the bacteria and can be used to find the torque and swimming speed in an indirect manner (though the dimensions of the flagellum are inferred and not directly measured in these experiments). The rotation rate of the cell body and flagellum were measured to be 19.6 Hz and 115 Hz respectively and the thrust force was found to be 0.57 pN. In the absence of the trap, swimming speeds of $22$ \SI{}{\micro\metre}$/s$ were measured using direct video microscopy. Using these values, Chattopadhyay et al.~\cite{chattopadhyay2006} estimated the motor torque value to be 500 pN~nm.  These values are close to that found in our simulations of an immobile bacterium close to the glass surface, 475 pN~nm (CM-I) and 440 pN~nm (CM-II). 
 In recent experiments, Drescher et al.~\cite{drescher2011} used flow field measurements to estimate  the force dipole  generated for a bacterium swimming at $22\pm 5$ \SI{}{\micro\metre}$/s$, and obtained a dipole consistent with a thrust force of 0.42 pN, close to the 0.57~pN found by Chattopadhyay et al.\cite{chattopadhyay2006}. Our numerical simulations (CM-I) for data sets A and B predict a thrust value of 0.28 pN and 0.49 pN for cells swimming at speeds 20.6 \SI{}{\micro\metre}$/s$ and 12.6 \SI{}{\micro\metre}$/s$ with their flagellum rotating at 127~Hz and 65.7~Hz respectively, in broad agreement. 

Shimogonya et al.\cite{shimogonya2015} performed tethered bead experiments using gold particles of diameter \SI{0.25}{\micro\metre}. They measured the precession of these particles and predicted a motor torque of the order of 700 pN~nm using boundary element method. 

Finally, the latest experiments using tethered beads method on \SI{1}{\micro\metre} diameter spheres performed by Van Oene et al.~\cite{van2017} suggest the motor torque to be 874 $\pm$ 206 pN~nm corresponding to a motor speed of 30.5 $\pm$ 6.9 Hz. These experimental measurements match our simulations in the case of a flagellar filament rotating very close to a wall. In the same experimenters, Van Oene et al.~\cite{van2017} were able to stop the rotation of the spherical magnetic beads by applying an external magnetic torque and found the stall motor torque be 444 $\pm$ 366 pN~nm, surprisingly much smaller than the prediction for the motor torque in the case of rotating tethered beads.

\subsection{Comparison with past simulations and theory}\label{sec:comparesim}

Using an analytical model based on resistive force theory, Darnton et al.~\cite{darnton2007}  estimated the motor torque of \textit{E.~coli} to be $370 \pm 100$ pN~nm, the lowest value among all studies.  

Mesoscale hydrodynamic simulations \cite{hu2015} have been performed for a bacterium with cell body length $2a=$ \SI{2.5}{\micro\metre}, width $2b=$\SI{0.9}{\micro\metre}, and a single flagellar filament with pitch $\lambda=$ \SI{2.2}{\micro\metre}, pitch angle $\psi=$ \ang{30}, cross-sectional radius $\rho=$ \SI{0.012}{\micro\metre} and 3 turns corresponding to an axial length of $\lambda=$ \SI{6.6}{\micro\metre} (instead of \SI{8.3}{\micro\metre} or \SI{10}{\micro\metre} in the experiments). These simulations predict a swimming speed of 14.5 \SI{}{\micro\metre}$/s$ and flagellum rotation frequency of 131 Hz resulting from  applying a motor torque equal to $\sim 1,200$ pN~nm. We note that the authors did not perform simulations of a stationary bacterium. However, by invoking linearity of Stokes flows, we can infer that the flagellum rotation frequency would be 111 Hz if a motor torque value  of $\sim 1,000$ pN~nm was  applied, allowing comparison with the data of Darnton et al.~\cite{darnton2007}. 

Simulations based on bead spring model \cite{kong2015} with three flagella also attempted to reproduce the experiments of Darnton et al.~\cite{darnton2007} with  flagellum dimensions which are not exactly the same but are close to the experiments. Specifically, the relevant dimensions used in these simulations are: cell body length $2a=$ \SI{2.5}{\micro\metre}, width $2b=$\SI{0.88}{\micro\metre}, pitch $\lambda=$ \SI{2.5}{\micro\metre} (instead of \SI{2.22}{\micro\metre} in the experiments), helical radius $R=$ \SI{0.5}{\micro\metre} (instead of \SI{0.4}{\micro\metre} in the experiments), cross-sectional radius $\rho=$ \SI{0.012}{\micro\metre} and axial length $\lambda=$ \SI{8.3}{\micro\metre}. The motor torque applied in these simulations is also 1,200 pN~nm. The cell, bundle rotation rate and swimming speed were found to be 26 Hz, 62 Hz and 24 \SI{}{\micro\metre}$/s$, respectively. On scaling the bundle rotation rate to the experimental value of Darnton et al.~\cite{darnton2007}, $111$ Hz, we obtain the motor torque value to be $\sim 2,150$ pN~nm.

%%%%%%%%%%%%%%%%%%%%%%%%%%%%
\section{Discussion}\label{sec:discussion}

In this paper, we have used two computational approaches based on the well established boundary integral equations and slender body theory valid for Stokes flow in order to compute the value of torque generated by the rotary motor of an {\it E.~coli}. We performed a number of tests to comprehensively validate the numerical simulations with analytical results presented in the appendices. The models are also validated with existing semi-analytical solutions of Higdon \cite{higdon1979b} relevant for micro-organisms swimming due to helical waves. We note that the model for the rotary motor used in the article is, in effect, a lumped model that represents the whole motor by a single rotation and associated torque. The hydrodynamics of the hook is ignored owing to its small size compared to the flagellar filament. Since, in  steady state, the elastic hook simply rotates with the rigid flagellum  it does not change the amount of torque transmitted to the flagellum. Equipped with our model, we first simulated the dynamics of {\it E.~coli} in the case where it is swimming due to a single rotating flagellar filament. In the former case, the agreement between our simulations the experiments of Darnton et al.~\cite{darnton2007} is quite remarkable even though the swimming cells in the experiments are propelled by multiple  interacting  flagellar filaments.

We next considered the situation where the  cell has a stationary cell body, an idealised situation to make a prediction for the value of the motor torque given that all parameters from the experiment of Darnton et al.~\cite{darnton2007} have been measured.  Values reported in the  literature for the motor torque span a wide range, mostly 500--1,200 pN~nm, and our computations are on the smaller end of that spectrum.  It is not clear at all what causes these numbers to differ so much from each other, even though many of the experiments employ the same techniques. As we have demonstrated here, the distance  between a rotating flagellar filament and a nearby surface is crucial to obtaining the correct values, as hydrodynamic friction depends critically on it, and suggests that perhaps hydrodynamic interactions with surface might have played an important role in some of the experimental investigations.

\appendix

\section{Discretisation method}\label{sec:discretisation}
\begin{figure}
\centering
\includegraphics[width=0.48\textwidth]{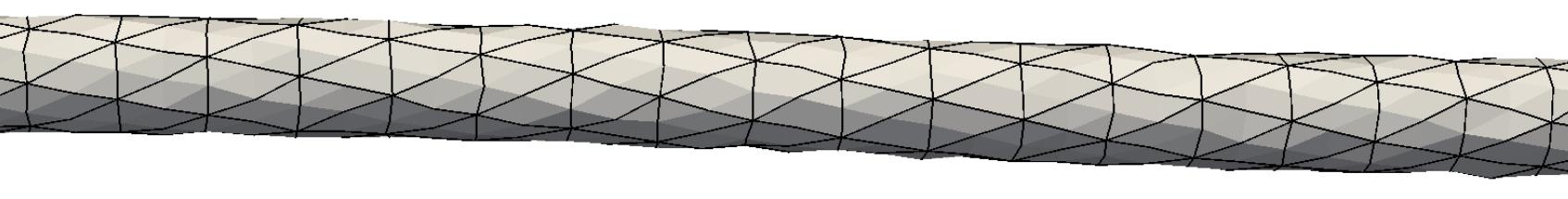}
\caption{Surface discretisation of a rigid helical object using 6-nodes curved elements. Evenly spaced circles are created along the centreline of the helix that are discretised into 12 points that either form a vertex or midpoints of the individual elements.}
\label{fig:helixmesh}
\end{figure}

In order to create a spherical mesh, we use the subroutine BEMLIB \cite{pozrikidis2002} starting from an icosahedron and successively subdividing the triangles. The transformation of the sphere to a prolate spheroid is done by simply rescaling the coordinates, $y=(b/a)y$ and $z=(b/a)z$. To create a helical surface, we first specify the number of points $N_{cl}$ along the centreline of the helix. Around each point, we then create a circle of radius $\rho$ having 12 points. These points either form a vertex or mid-points of the 6-nodes triangles. Each consecutive circle is shifted by $\pi/24$ so that we have approximately uniform isosceles triangles rather than right-angled triangles. A section of a discretised helix is shown in Fig.~\ref{fig:helixmesh}. We attach 2 hemispheres of radius $\rho$ on both ends of the discretised helix to remove sharp corners.

\section{Green's function for a stokeslet near a wall} \label{sec:blakelets}

Let us consider a stokeslet placed above a wall at $z=0$ at a distance $h$ such that its location is $(y_1,y_2,h)$. The image singularities are then accordingly located below the wall at $(y_1,y_2,-h)$. The Green's function \cite{blake1974} due to the stokeslet at an evaluation point $(x_1,x_2,x_3)$ is, 
\begin{equation}
\begin{split}
G_{ij}^w = &\frac{\delta_{ij} + \hat{r}_i\hat{r}_j}{r} -\frac{\delta_{ij} + \hat{R}_i\hat{R}_j}{R} \\
& + 2h \Delta_{jk}\frac{\partial}{\partial R_k}\left(\frac{h\hat{R}_i}{R^2} - \frac{\delta_{i3} + \hat{R}_i\hat{R}_3 }{R}\right),
\end{split}
\end{equation}
where the vector pointing from the stokeslet location to the evaluation point is $r_i=(x_1-y_1,x_2-y_2,x_3-h)$, the vector pointing from the image location to the evaluation point is $R_i=(x_1-y_1,x_2-y_2,x_3+h)$. The matrix $\Delta_{jk}$ takes the value of 1 when $j=k=1,2$, -1 when $j=k=3$ and 0 for every other combination.

\section{Interaction between two spheres}\label{sec:spheres}

\begin{figure}[b]
\centering
\includegraphics[width=0.48\textwidth]{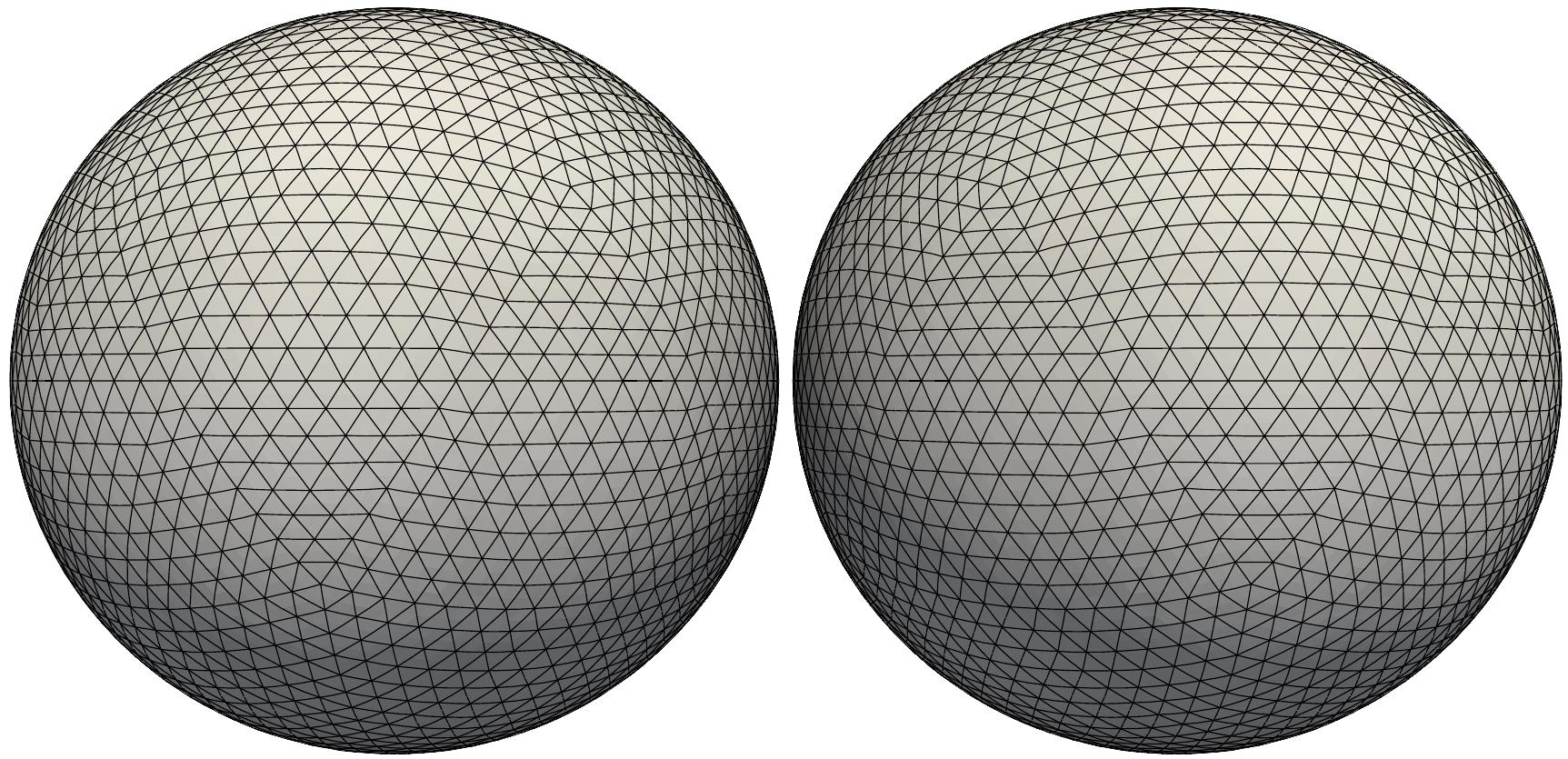}
\caption{Boundary element mesh ($N_b=5120$) of two spheres separated by a distance $d$ approaching each other with velocities, $\bm{U}$, along the line connecting the centres.}
\label{fig:spheres}
\end{figure}

\begin{figure}
\centering
\includegraphics[width=0.48\textwidth]{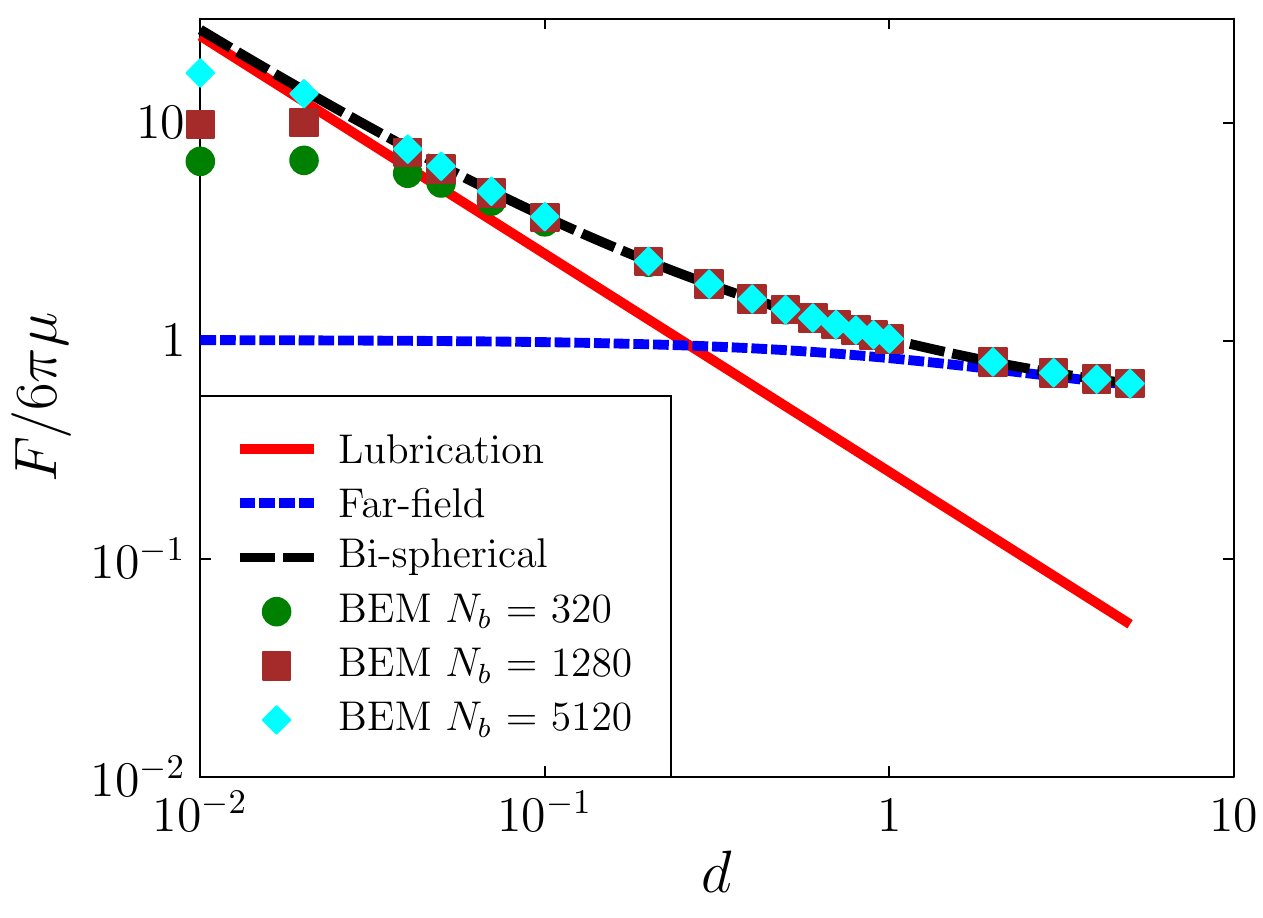}
\caption{Comparison of hydrodynamic force acting on 2 spheres approaching each other at a relative velocity of 1 along the line joining their centres obtained by boundary element method and theoretical calculations.}
\label{fig:lubrication}
\end{figure}

In order to validate our numerical algorithm for two solid interacting bodies in Stokes flow, we analyse the hydrodynamic force acting on 2 spheres approaching each other, separated by a distance $d$. Fig.~\ref{fig:spheres} shows two discretised spheres, each made with 5120 elements. The line joining their centres is along the $x$ axis. We essentially solve equations \eqref{eq:bemslp-model-1} to find the tractions on the 2 spheres moving with a prescribed velocity of $\bm{U}= \pm 0.5 \bm{\hat{i}}$, so that the relative speed of approach is 1. We compare the force computed by our numerical method for 3 different discretisation levels, $N_b =$ 320, 1280 and 5120 with that obtained using lubrication theory \cite{kim2013} valid for small distances, far-field asymptotic solution valid for large distances and exact solutions based on bispherical coordinates, see Fig.~\ref{fig:lubrication}. Our numerical method based on boundary element method finds excellent agreement with the theoretical predictions. Note that as the distance between the spheres $d$ decreases, finer discretisation of the spheres near the closest point becomes necessary to resolve the flow field accurately. In order to validate the force-free and torque-free bacterium model, we can think of sphere 2 as a rotating flagella and implement computational model I. On imposing a relative velocity of $\bm{\Omega}_m = \bm{\hat{i}}$, we found the body rotation rate to be $\bm{\Omega}_b = -0.5\bm{\hat{i}}$, as expected theoretically. The swimming velocity $\bm{U}$ is found to be exactly zero as neither of the bodies are chiral. This serves as another validation for computational model I.

\section{Rotating and translating slender cylinder}\label{sec:cylinder}

\begin{figure}
\centering
\includegraphics[width=0.48\textwidth]{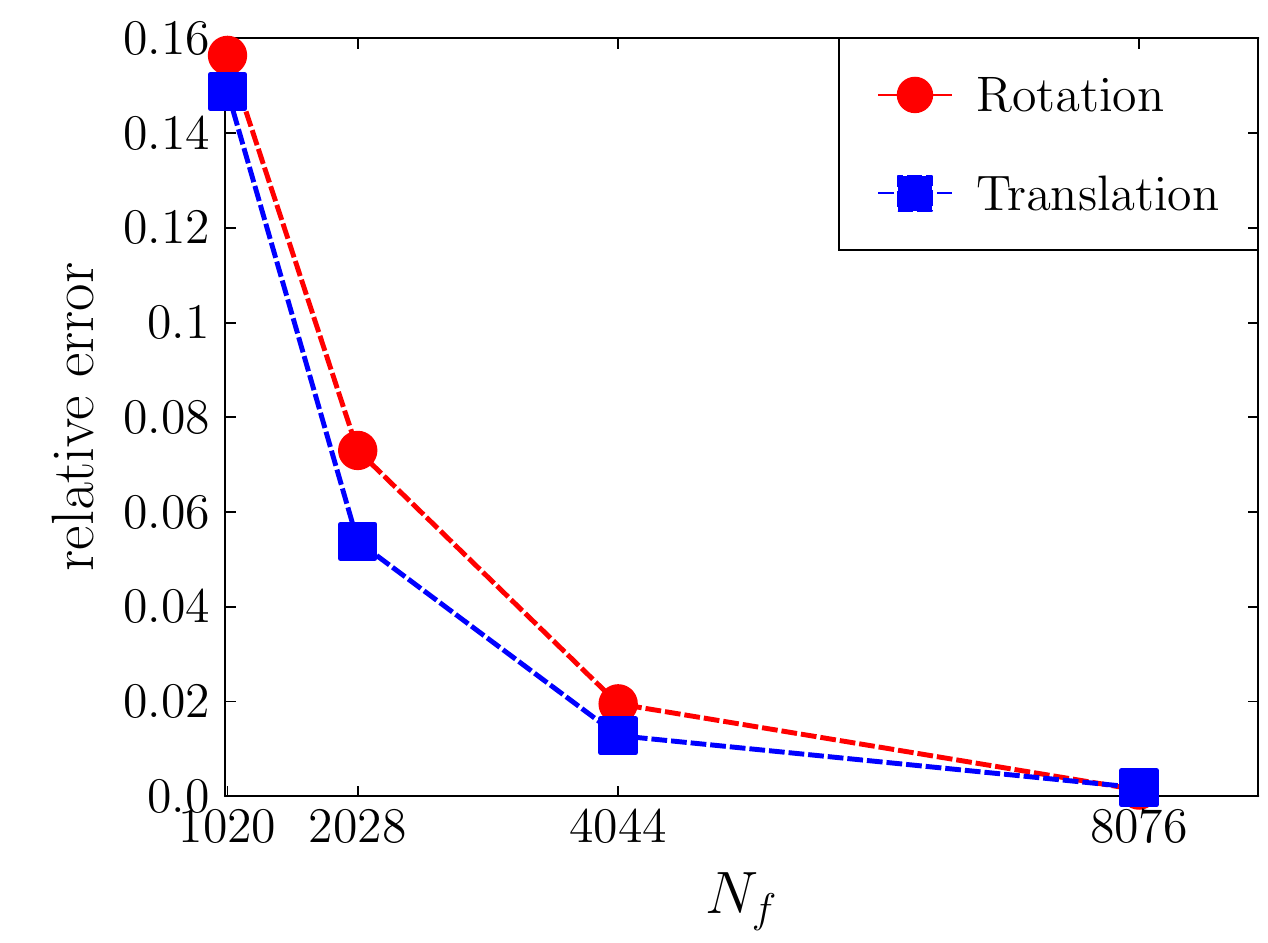}
\caption{Relative error in the hydrodynamic torque and force per unit length acting on a slender cylinder rotating in an infinite fluid medium and translating perpendicular to its axis next to a plane wall, respectively, computed using boundary element methods for 4 levels of discretisation $N_f =$ 1020, 2028, 4044 and 8076.}
\label{fig:cylinder}
\end{figure}

As there are no exact solutions to problems of helices in viscous flows, we test our numerical method for slender cylinders rotating in an infinite fluid medium and translating next to a plane wall. The length of the cylinder is $L=7$ while the cross-sectional radius is $\rho=0.01$, so that the aspect ratio is $\varepsilon = 0.0014$, comparable to that of a bacterial flagella. An infinitely long cylinder rotating in a viscous fluid experiences a viscous torque per unit length \cite{kim2013} $\bm{T}_h= -4\pi \mu \rho^2 \bm{\Omega}$ with the angular velocity $\bm{\Omega}$ pointing along the axis of the cylinder. Using boundary element method, we obtain the hydrodynamic torque per unit length acting on the discretised rotating cylinder. 
Fig.~\ref{fig:cylinder} shows the relative error between the numerical simulations and analytical results for 4 different levels of discretisation, $N_f =$ 1020, 2028, 4044 and 8076.

In order to validate our results that include wall effects, we consider a slender cylinder translating perpendicular to its long axis close to a wall such that the cylinder centreline is at a distance $d=0.02$ from the wall. The dimensions of the cylinder are the same as consider before. The translating cylinder experiences a hydrodynamic force per unit length \cite{jeffrey1981} $\bm{F}_h = -4 \pi \mu \bm{U}/\alpha$, where $\alpha= \log{[(d/\rho) + \sqrt{(d/\rho)^2-1}]}$. Fig.~\ref{fig:cylinder} shows the relative error between the numerical simulations and analytical results for 4 different levels of discretisation, $N_f =$ 1020, 2028, 4044 and 8076. In both these tests, we find excellent agreement between numerics and theory and the accuracy increases as we increase the grid resolution.

%%%%%%%%%%%%%%%%%%%%%%%%%%%%
\section*{Acknowledgements}
We thank Howard Berg, Richard Berry, Maciej Lisicki, John Lister and Thomas Powers  for helpful discussions and Sebastien Michelin for providing the MATLAB code for the exact solution of the force acting on two   spheres presented in the appendix~\ref{sec:spheres}. This project has received funding from the European Research Council (ERC) under the European Union's Horizon 2020 research and innovation programme  (grant agreement 682754 to EL).

\bibliography{papers} 
\bibliographystyle{unsrt}

\end{document}